# A Study of N-Acetyl aspartic acid/Creatine Ratio in the White Matter of HIV Positive Patients and Its Application

Vishal Midya[1], Ulhaas Shankar Chakraborty[2]

[1]Indian Statistical Institute, Delhi, India
[2] Seth G.S. Medical College & KEM Hospital

A total of 50 patients were enrolled in the study, and MRI brain with MR spectroscopy was done. Out of these, 35 were HIV positive patients (cases). 15 patients with HIV negative status who also underwent MRI for unrelated pathologies were enrolled as controls. They also underwent MRI brain with MR spectroscopy. Tuberculosis was the most common neurologic disease found in the HIV positive group, consisting of 9 patients. Two patients had healed granulomas. Seven of these patients had tuberculous meningitis amongst which a further 2 had vasculitic infarcts. Another two in this group had tuberculomas. Prominent lipid lactate peaks on the spectra obtained from the lesions confirmed the diagnosis of tuberculoma. PML was seen in 6 patients. NAA/Cr was found to be reduced in all the patients, and in fact the value was further reduced compared to the HIV positive group as a whole. Raised choline and myoInositol peaks were also found in all the patients. Toxoplasmosis was seen in 4 patients. MR Spectroscopy showed lipid – lactate peaks confirming the diagnosis. A single patient with brain abscess was seen where also the spectroscopic data from the lesion confirmed the diagnosis with a lactate peak as well as amino acid resonances at characteristic locations. 2 patients had HIV encephalopathy on the imaging study. Their spectra also revealed lowered NAA peaks along with raised choline peaks. 2 patients with cryptococcosis showed characteristic imaging finding of enlarged Virchow Robin (perivascular) spaces. They revealed elevated choline peaks in addition to reduced NAA. No trehalose peaks were identified. MR Spectroscopy was also done in 4 patients of vascular thrombosis, two arterial and two venous, where the MR Spectroscopic findings were not specific. In 7 patients no focal lesions could be identified on conventional MRI sequences however on MR spectroscopic analysis NAA was found to be reduced in comparison to the controls. The values of NAA/Cr were determined after duly processing the spectroscopic data from both cases and controls. Each group (Cases and controls) were divided on the basis of age into two age groups: Lesser than or equal to 40 years, and greater than 40 years. In all three groups the values of the mean NAA/Cr ratio was significantly (p-value less than 0.05) reduced in comparison to controls. An ancillary finding was the reduction of NAA/Cr further in cases of PML. Combined use of both the conventional and advanced MRI sequences is advisable as spectroscopy helps in confirming the diagnosis of opportunistic infection of the CNS in HIV positive patients. NAA/Cr ratio is reduced in HIV positive patients and is a marker for HIV infection of the brain even in the absence of imaging findings of HIV encephalopathy or when the patient is symptomatic due to neurologic disease of other etiologies.

# INTRODUCTION

The Human Immunodeficiency Virus (HIV) is the retrovirus responsible for AIDS. There are two subtypes of HIV, HIV-1 and HIV-2. HIV-1 is responsible for the majority of the AIDS cases throughout the world, whereas HIV-2 is rarely to be found outside West Africa[1].

On June 5, 1981, the Morbidity and Mortality Weekly Report (MMWR), published by the Centers for Disease Control and Prevention (CDC), reported *Pneumocystis* pneumonia in 5 homosexual men. It documented for the first time in history what became known as acquired immunodeficiency syndrome (AIDS). Within a month, additional diagnoses of opportunistic infections in homosexual men from New York City and California were reported. These articles were harbingers for what became one of history's worst pandemics, with more than 60 million infections, more than 30 million deaths, and 30 years later, with no definitive cure[1].

The Human Immunodeficiency Virus (HIV) is neurovirulent[2], with up to half of all patients with HIV exhibiting neurologic disease, in spite of the establishment of effective antiretroviral therapy[3]. Thus it is a major public health problem.

The study of Central Nervous System disease in HIV infection (CNS-D) in HIV infection is complicated by the fact that in addition to the direct effects of HIV on the brain, multiple other disease processes including opportunistic infection, neoplasm and immune phenomena contribute significantly to the disease burden, all with overlapping imaging features[4,5]. It is significant to note that the latest figures for adult prevalence of HIV infection in India from NACO (National AIDS control Organization) available are 0.31%. India has the third largest population of people living with HIV/AIDS in the world. While the effect of opportunistic infections have reduced since the advent of HAART (Highly active antiretroviral therapy), as well as adequate chemoprophylaxis against opportunistic infections, given the proportion of patients with HIV who have neurologic disease, the morbidity of HIV related brain injury thus remains an important health problem.

MRI is a more sensitive tool in the workup of HIV positive patients with CNS-D. It helps in the diagnosis of specific disease, monitoring of effects of therapy, to guide stereotactic biopsy if required, and also to prognosticate cases[5,6]. Although MR imaging is currently the most sensitive imaging modality in depicting brain alterations in HIV positive individuals, conventional MR sequences may not be able to demonstrate any alterations in the early stages of HIV infection[7]. Of the advanced neuroimaging modalities, Magnetic resonance spectroscopy is especially useful as it non-invasively provides essential information for understanding the underlying causes of CNS injury in the HIV-infected patient. By quantifying the surrogate markers of *N*-acetylaspartate (NAA), creatine (Cr), choline (Cho), lactate and myo-inositol (mI), MRS offers insight into the neuronal integrity, cell membrane synthesis and turnover, macrophage infiltration, inflammation status, and levels of microglial activation and gliosis within the sampled CNS tissue[8]. Many studies have also shown metabolite abnormalities early in the pathogenesis of HIV infection, earlier than detectable with conventional MR sequences[8].

HIV encephalopathy is the most common neurologic complication of HIV infection[7,9]. The most common opportunistic infections include Toxoplasmosis, Cryptococcosis, PML, Mycobacterial infection, CMV and pyogenic infection[10]. Neoplastic diseases associated with HIV infection include Primary CNS lymphoma and metastatic Kaposi's Sarcoma[9]. Non-infective and non-neoplastic neurologic complications of HIV infections include demyelinating disorders, vascular disorders, Parkinson's disease, and spontaneous subdural bleeding[11].

# REVIEW OF LITERATURE

## 1. HISTORY

In view of the growing importance and near universal applicability of MRI in the practice of medicine, Paul Lauterbur of the University of Illinois at Urbana-Champaign and Sir Peter Mansfield of the University of Nottingham were awarded the 2003 Nobel Prize in Physiology or Medicine for their "discoveries concerning magnetic resonance imaging". The Nobel citation noted Lauterbur's insight of using magnetic field gradients to determine spatial localization, a breakthrough discovery that allowed rapid acquisition of 2D images. Mansfield was credited with introducing the mathematical formalism and developing techniques for efficient gradient utilization and fast imaging.

In the 1950s, Herman Carr reported on the creation of a one-dimensional MR image[12]. Paul Lauterbur expanded on Carr's technique and developed a way to generate the first MRI images, in 2D and 3D, using gradients. In 1973, Lauterbur published the first nuclear magnetic resonance image and the first cross-sectional image of a living mouse was published in January 1974. Nuclear magnetic resonance imaging is a relatively new technology first developed at the University of Nottingham, England. Peter Mansfield, a physicist and professor at the university, then developed a mathematical technique that would allow scans to take seconds rather than hours and produce clearer images than Lauterbur had.

In a 1971 paper in the journal *Science*[13], Dr. Raymond Damadian, an Armenian- American physician, scientist, and professor at the Downstate Medical Center State University of New York, was responsible for one of the first reported clinical uses of the MR Scanning. He reported that tumors and normal tissue can be distinguished in vivo by nuclear magnetic resonance ("NMR"). He suggested that these differences could be used to diagnose cancer, though later research would find that these differences, while real, are too variable for diagnostic purposes. Damadian's initial methods were flawed for practical use, relying on a point-by-point scan of the entire body and using relaxation rates, which turned out to not be an effective indicator of cancerous tissue.

While researching the analytical properties of magnetic resonance, Damadian was responsible for the creation of the world's first magnetic resonance imaging machine in 1972. He filed the first patent for an MRI machine, on March 17, 1972, which was later issued to him on February 5, 1974. As the National Science Foundation notes, "The patent included the idea of using NMR to 'scan' the human body to locate cancerous tissue. However, it did not describe a method for generating pictures from such a scan or precisely how such a scan might be done‖. Damadian along with Larry Minkoff and Michael Goldsmith subsequently went on to perform the first MRI body scan of a human being on July 3, 1977. These studies performed on humans were published in 1977. In recording the history of MRI, Mattson and Simon (1996) credit Damadian with describing the concept of whole-body NMR scanning, as well as discovering the NMR tissue relaxation differences that made this feasible Nuclear magnetic resonance (NMR) spectroscopy in the solid and liquid states were first described independently by Felix Bloch and Edward Mills Purcell independently in 1946 and they shared the Nobel Prize for their discovery in 1952. It is generally known and has been used extensively as an analytical method in chemistry to identify molecules and to determine their biophysical characteristics. The clinical use of in vivo magnetic resonance spectroscopy (MRS) has been limited for a long time, mainly due to its low sensitivity. However, with the advent of clinical MR systems with higher magnetic field strengths, the development of better coils, and the design of optimized radio-frequency pulses, sensitivity has been considerably improved. Therefore, in vivo MRS has become a technique that is routinely used more and more in clinical radiologic practice[14]. There are about 20,000 metabolites in the human body[15]. MRS is an ideal technique to measure changes in metabolite levels non-invasively, and thus the technique is very suitable in the diagnostics of metabolic disorders. However, it should be noted that only small metabolites with sufficiently high concentration can be observed by MRS. General Pattern changes in combination with other clinical and biochemical information may help to find the correct diagnosis. As time passes and clinical experience with MRS increases, a wider range of disorders can be identified with pattern changes in MRS spectra.

## 2. MRI SEQUENCES

**$T_1$-weighted MRI**

$T_1$-weighted scans are a standard basic scan, in particular differentiating fat from water - with water darker and fat brighter use a gradient echo (GRE) sequence, with short $T_E$ and short $T_R$. This is one of the basic types of MR contrast and is a commonly run clinical scan. The $T_1$ weighting can be increased (improving contrast) with the use of an inversion pulse as in an MP-RAGE sequence. Due to the short repetition time ($T_R$) this scan can be run very fast allowing the collection of high resolution 3D datasets. A $T_1$ shortening gadolinium contrast agent is also commonly used, with a $T_1$ scan being collected before and after administration of contrast agent to compare the difference. In the brain $T_1$-weighted scans provide good gray matter/white matter contrast; in other words, $T_1$-weighted images highlight fat deposition.

**$T_2$-weighted MRI**

$T_2$-weighted scans are another basic type. Like the $T_1$-weighted scan, fat is differentiated from water - but in this case fat shows darker, and water lighter. For example, in the case of cerebral and spinal study, the CSF (cerebrospinal fluid) will be lighter in $T_2$-weighted images. These scans are therefore particularly well suited to imaging edema, with long $T_E$ and long $T_R$. Because the spin echo sequence isless susceptible to inhomogeneities in the magnetic field, these images have long been a clinical workhorse.

T*2-weighted MRI scans use a gradient echo (GRE) sequence, with long $T_E$ and long $T_R$. The gradient echo sequence used does not have the extra refocusing pulse used in spin echo so it is subject to additional losses above the normal $T_2$ decay (referred to as $T_2'$), these taken together are called T2*. This also makes it more prone to susceptibility losses at air and tissue boundaries, but can increase contrast for certain types of tissue, such as venous blood.

**Proton density weighted MRI**

Spi density, also called proton density, weighted scans try to have no contrast from either $T_2$ or $T_1$ decay, the only signal change coming from differences in the amount of available spins (hydrogen nuclei in water). It uses a spin echo or sometimes a gradient echo sequence, with short $T_E$ and long $T_R$.

**Diffusion Weighted Imaging:**

Diffusion imaging (DWI) is a modification of regular MRI techniques, and is an approach which utilizes the measurement of Brownian motion of molecules. Regular MRI acquisition utilizes the behavior of protons in water to generate contrast between clinically relevant features of a particular subject. The versatile nature of MRI is due to this capability of producing contrast, called weighting. In a typical $T_1$-weighted image, water molecules in a sample are excited with the imposition of a strong magnetic field. This causes many of the protons in water molecules to
Process simultaneously, producing signals in MRI. In $T_2$-weighted images, contrast is produced by measuring the loss of coherence or synchrony between the water protons. When water is in an environment where it can freely tumble, relaxation tends to take longer. In certain clinical situations, this can generate contrast between an area of pathology and the surrounding healthy tissue.
In diffusion-weighted images, instead of a homogeneous magnetic field, the homogeneity is varied linearly by a pulsed field gradient. Since precession is proportional to the magnet strength, the protons begin to process at different rates, resulting in dispersion of the phase and signal loss. Another gradient pulse is applied in the same direction but with opposite magnitude to refocus or rephase the spins. The refocusing will not be perfect for protons that have moved during the time interval between the pulses, and the signal measured by the MRI machine is reduced. This reduction in signal due to the application of the pulse gradient can be related to the amount of diffusion that is occurring through the following equation:

$$\frac{S}{S_0} = e^{-\gamma^2 G^2 \delta^2 (\Delta - \delta/3) D} = e^{-bD}$$

where $S_0$ is the signal intensity without the diffusion weighting, $S$ is the signal with the gradient, $\gamma$ is the gyromagnetic ratio, $G$ is the strength of the gradient pulse, $\delta$ is the duration of the pulse, is the time between the two pulses, and finally, $D$ is the diffusion-coefficient.

By rearranging the formula to isolate the diffusion-coefficient, it is possible to obtain an idea of the properties of diffusion occurring within a particular voxel (volume picture element). These values, called apparent diffusion coefficients (ADC) can then be mapped as an image, using *diffusion* as the contrast. The first successful clinical application of DWI was in imaging the brain following stroke in adults. Areas which were injured during a stroke showed up "darker" on an ADC map compared to healthy tissue. At about the same time as it became evident to researchers that DWI could be used to assess the severity of injury in adult stroke patients, they also noticed that ADC values varied depending on which way the pulse gradient was applied. This orientation-dependent contrast is generated by diffusion anisotropy, meaning that the diffusion in parts of the brain has directionality. This may be useful for determining structures in the brain which could restrict the flow of water in one direction, such as the myelinated axons of nerve cells (which is affected by multiple sclerosis). However, in imaging the brain following a stroke, it may actually prevent the injury from being seen. To compensate for this, it is necessary to apply a mathematical operator, called a tensor, to fully characterize the motion of water in all directions.

## 3. MR SPECTROSCOPY (MRS)

Magnetic resonance spectroscopy (MRS) is used to measure the levels of different metabolites in body tissues. MR spectra may be obtained from different nuclei, including phosphorus ($^{31}$P), fluorine ($^{19}$F), carbon ($^{13}$C) and sodium ($^{23}$Na). The ones mostly used for clinical MRS are protons ($^{1}$H) [16]. The brain is ideally imaged with Proton Magnetic Resonance Spectroscopy (H-MRS) because of its near lack of motion (this prevents MRS from being used in the abdomen and thorax without very sophisticated motion-reduction techniques). Protons ($^{1}$H) are the most used nuclei for clinical applications in the human brain mainly because of its high sensitivity and abundance. The proton MR spectrum is altered in almost all neurological disorders. The MR signal produces a spectrum of resonances that correspond to different molecular arrangements of the isotope being "excited". This signature is used to diagnose certain metabolic disorders, especially those affecting the brain, and to provide information on tumor metabolism.

Magnetic resonance spectroscopic imaging (MRSI) combines both spectroscopic and imaging methods to produce spatially localized spectra from within the sample or patient. The spatial resolution is much lower (limited by the available SNR), but the spectra in each voxel contains information about many metabolites. Because the available signal is used to encode spatial and spectral information, MRSI requires high SNR achievable only at higher field strengths.

**Basic principle:**

H-MRS is based on the chemical shift properties of the atom. When a tissue is exposed to an external magnetic field, its nuclei will resonate at a frequency (f) that is given by the Larmor equation:

$$f = \gamma B_0$$

Since the gyromagnetic ratio ($\gamma$) is a constant of each nuclear species, the spin frequency of a certain nuclei (f) depends on the external magnetic field ($B_0$) and the local microenvironment. The electric shell interactions of these nuclei with the surrounding molecules cause a change in the local magnetic field leading to a change on the spin frequency of the atom (a phenomenon called chemical shift). The value of this difference in resonance frequency gives information about the molecular group carrying 1H and is expressed in parts per million (ppm). The resonant frequency, energy of the absorption and the intensity of the signal are proportional to the strength of the magnetic field.

Chemical shift**:** Depending on their local chemical environment, different nuclei in a molecule absorb at slightly different frequencies. Since this resonant frequency is directly proportional to the strength of the magnetic field, the shift is converted into a field-independent dimensionless value known as the chemical shift. The chemical shift is reported as a relative measure from some reference resonance frequency. (For the nuclei $^{1}$H, $^{13}$C, and $^{29}$Si, TMS (tetramethylsilane) is commonly used as a reference.) This difference between the frequency of the signal and the frequency of the reference is divided by frequency of the reference signal to give the chemical shift. The frequency shifts are extremely small in comparison to the fundamental NMR frequency. A typical frequency shift might be 100 Hz, compared to a fundamental NMR frequency of 100 MHz, so the chemical shift is generally expressed in parts per million (ppm). To detect such small frequency differences the applied magnetic field must be constant throughout the sample volume. High resolution NMR spectrometers use shims to adjust the homogeneity of the magnetic field to parts per billion (ppb) in a volume of a few cubic centimeters. In general, chemical shifts for protons are highly predictable since the shifts are primarily determined by simpler shielding effects (electron density), but the chemical shifts for many heavier nuclei are more strongly influenced by other factors including excited states ("paramagnetic" contribution to shielding tensor).

The chemical shift provides information about the structure of the molecule. The conversion of the raw data to this information is called *assigning* the spectrum. Software allows analysis of the size of peaks to understand how many protons give rise to the peak. This is known as integration—a mathematical process which calculates the area under a curve. The analyst must integrate the peak and not measure its height because the peaks also have width—and thus its size is dependent on its area not its height. However, it should be mentioned that the number of protons, or any other observed nucleus, is only proportional to the intensity, or the integral, of the NMR signal.

J-coupling: Some of the most useful information for structure determination in a one-dimensional NMR spectrum comes from J-coupling or scalar coupling (a special case of spin-spin coupling) between NMR active nuclei. This coupling arises from the interaction of different spin states through the chemical bonds of a molecule and results in the splitting of NMR signals. These splitting patterns can be complex or simple and, likewise, can be straightforwardly interpretable or deceptive. This coupling provides detailed insight into the spacial relationship of atoms in a molecule. Coupling combined with the chemical shift (and the integration for protons) tells us not only about the chemical environment of the nuclei, but also the number of neighboring NMR active nuclei within the molecule.

Magnetic inequivalence: More subtle effects can occur if chemically equivalent spins (i.e., nuclei related by symmetry and so having the same NMR frequency) have different coupling relationships to external spins. Spins that are chemically equivalent but are not indistinguishable (based on their coupling relationships) are termed magnetically in equivalent.

**Techniques:**

The H-MRS acquisition usually starts with anatomical images, which are used to select a volume of interest (VOI), where the spectrum will be acquired. For the spectrum acquisition, different techniques may be used including single- and multi-voxel imaging using both long and short echo times (TE). Each technique has advantages and disadvantages and choosing the right one for a specific purpose is important to improve the quality of the results.

Single Voxel Spectroscopy: In single voxel spectroscopy (SVS) the signal is obtained from a voxel previously selected. This voxel is acquired from a combination of slice-selective excitations in three dimensions in space, achieved when a RF pulse is applied while a field gradient is switched on. It results in three orthogonal planes and their intersection corresponds to the voxel of interest. There are two main techniques: point resolved spectroscopy (PRESS) and stimulated echo acquisition mode (STEAM).

In the PRESS sequence, the spectrum is acquired using one 90 degree pulse followed by two 180 degree pulses. Each of them is applied at the same time as a different field gradient. Thus, the signal emitted by the VOI is a spin echo. The first 180 degree pulse is applied after a time ½ TE from the first pulse (90 degree pulse) and the second 180 degree is applied after a time ½ TE + TE. The signal occurs after a time 2TE (Fig. 3). To restrict the acquired sign to the VOI selected, spoiler gradients are needed. Spoiler gradients dephase the nuclei outside the VOI and reduce their signal.

In the STEAM sequence all three pulses applied are 90 degree pulses. As in PRESS, they are all simultaneous with different field gradients. After a time ½ TE from the first pulse, a second 90 degree pulse is applied. The time elapsed between the second and the third is conventionally called ―mixing time‖ (MT) and is shorter than ½ TE. The signal is finally achieved after a time TE + MT from the first pulse. Thus, the total time for STEAM technique is shorter than PRESS. Spoiler gradients are also needed to reduce signal from regions outside the VOI.

Multivoxel spectroscopy: Multivoxel spectroscopy has the advantage of being able to collect pieces of data from multiple voxels of interest simultaneously, enabling the understanding of the spatial heterogeneity of the distribution of metabolites. The method is known as Chemical Shift Imaging (CSI) or 2D Magnetic Resonance Spectroscopy Imaging (MRSI). The same sequences used for SVS are used for the signal acquisition in MRSI (STEAM or PRESS). The main difference between 2D MRSI and SVS is that, after the RF pulse, phase encoding gradients are used in two dimensions to sample the k-space. The results are expressed in the form of a spectroscopy grid, with spectra associated with each voxel in a grid.

Water Suppression - Techniques and Utility: MRS-visible brain metabolites have a low concentration in brain tissues. Water is the most abundant and thus its signal in MRS spectrum is much higher than that of other metabolites (the signal of water is 100.000 times greater than that of other metabolites). To avoid this high peak from water to be superimpose on the signal of other brain metabolites, water suppression techniques are needed. The most commonly used technique is chemical shift selective water suppression (CHESS) which pre-saturates water signal using frequency selective 90o pulses before the localizing pulse sequence. Other techniques sometimes used are VAriable Pulse power and Optimized Relaxation Delays (VAPOR) and Water suppression Enhanced through T1 effects (WET).

Acquisition parameters - Long and Short TE: It is important to note that MRS can be obtained using different TEs that result in distinct spectra. Short TE refers to a study in which it varies from 20 to 40 ms. It has a higher SNR and less signal loss due to T2 and T1 weighting than long TE. These short TE properties result in a spectrum with more metabolites peaks, such as myoinositol and glutamine-glutamate whichare not detected with long TE. However the baseline is more complicated due to the increased lipid and macromolecular background signals.
MRS spectra may also be obtained with long TEs, from 135 to 288 ms. Long TEs have a worse SNR, but they have simpler spectra due to suppression of some signals. Thus, the spectra are less noisy but have a limited number of sharp resonances. Hence interpretation and analysis of the spectra are simpler. Also Lactate and alanine peaks are inverted, helping to differentiate them from the lipid / macromolecules peaks. With even longer TRs the SNR improves further. However this improvement comes at the expense of longer scan times. As the main limitation of MRS is the long scan times, this is undesirable and all clinical applications of under the curve formed by the metabolite resonance peak is proportional to the concentration of MRS have a TE less than 3 sec.

Quantification methods: As we know the area that specific metabolite. By the mathematical process of integration, we can quantify metabolites. Both absolute and relative concentrations can be obtained. However though absolute concentrations are more accurate, they are often not reproducible and errors in measurement are much higher. An easier and better method is to use metabolite ratios. Creatinine is a metabolite the concentration of which does not often changes during disease processes, and is hence used as an internal standard. Though the assumption that the concentration of creatinine remains constant over time is not always true, it is almost always true. Hence ratios are widely used as the best spectroscopic quantification method in following the course of neurologic diseases.

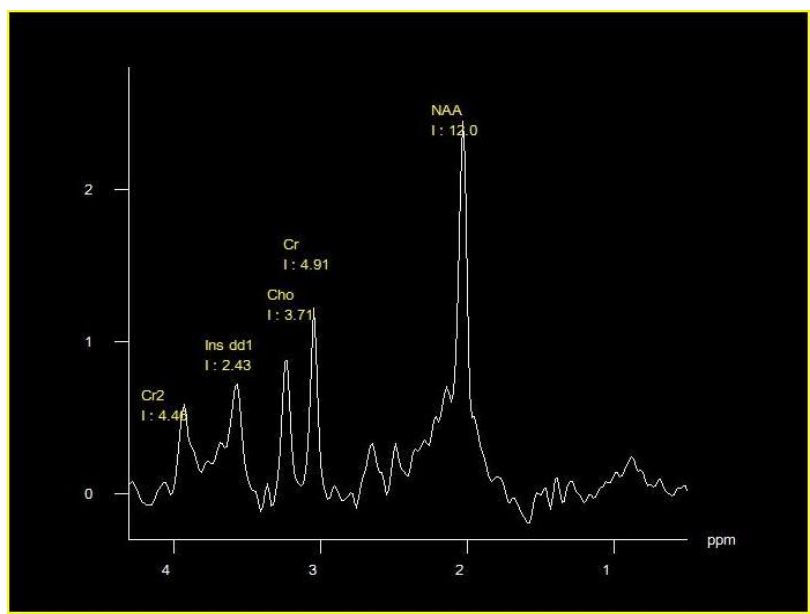

Fig: The above figure shows a spectrum obtained from the centrum semiovale of a normal person. The most prominent peak is N-acetyl aspartate at 2.02ppm. Following from right to left: creatine (Cr) at 3.02 (and 3.9), choline (Cho) at 3.2 and myoinositol at 3.56 ppm.

## 4. PATHOGENESIS

HIV is a retrovirus that has a unique replication cycle. The retrovirus enters a target cell by binding to a specific cell-surface receptor; once the virus is internalized, its RNA is released from the nucleocapsid and is reverse-transcribed into proviral DNA. The provirus is inserted into the genome and then transcribed into RNA; the RNA is translated; and new viruses assemble and are extruded from the cell membrane by budding[17]. The hallmark of HIV disease is a profound immunodeficiency resulting primarily from a progressive quantitative and qualitative deficiency of the subset of T lymphocytes referred to as *helper T cells (*CD4+*)* occurring in a setting of polyclonal immune activation. Patients with CD4+ T cell levels below certain thresholds are at high risk of developing a variety of opportunistic diseases.

Apart from opportunistic infections, HIV can cause damage to the CNS as a direct result of viral infection of the CNS macrophages or glial cells or secondary to the release of neurotoxins and cytokines. In addition genetic variables may influence susceptibility of patients to CNS disease[17]. Virtually all patients with HIV infection have some degree of nervous system involvement with the virus. This is evidenced by the fact that CSF findings are abnormal in ~90% of patients, even during the asymptomatic phase of HIV infection. CSF abnormalities include pleocytosis (50– 65% of patients), detection of viral RNA (~75%), elevated CSF protein (35%), and evidence of intrathecal synthesis of anti-HIV antibodies (90%).

Pathologically, characteristic changes are seen in the brains of persons with HIV which can exist singly or overlap. The most common finding is cerebral atrophy with resultant ventricular enlargement and widened sulci. Extensive inflammatory changes are seen in both the deep white and deep gray matter. Distinct pockets of inflammatory cells, such as microglia, macrophages, and multinucleated giant cells formed by fusion of microglia and macrophages, are found in white and deep grey matter as well as in the cortex. In addition to neurons, glial cells also harbour HIV. HIV leukoencephalopathy is characterized by diffuse damage to white matter or myelin attenuation, astrogliosis, macrophages, and multinucleated giant cells without distinct pockets of inflammation[18]. Electron microscopy shows the multinucleate giant cells to be packed with retrovirus particles. The presence of these viral loaded cells is strongly associated with AIDS Dementia complex[7]. HIV related poliodystrophy is a diffuse atrophy of the gray matter. Reactive astrogliosis and microglial activation in cerebral gray matter may possibly be associated with neuron loss or dendritic damage. Severe cases may also evidence spongiform changes. Demyelination apparent on imaging is a comparatively late finding[7,18].

Vascular calcification of the basal ganglia and possibly the centrum semiovale is often present in pediatric AIDS. In fact, it may be found either with or without notable HIV-1 encephalitis. Areas of myelin pallor in young children may be difficult to distinguish because of their incomplete myelination. In contrast, inflammatory changes are more common in the spinal cord in pediatric patients, whereas vacuolar myelopathy is seen less often. However, corticospinal tract degeneration may also be found in pediatric HIV-1 infection[18].

## 5. PATHOLOGIC, MR IMAGING, AND MR SPECTROSCOPIC FEATURES

**Toxoplasmosis:**

Toxoplasmosis is the most frequent cause of intracranial mass lesion in the HIV positive population[19]. It is an obligate intracellular protozoan. Transmission of infection occurs commonly through the ingestion of food or water contaminated with cat feces, or transplacentally. Infection is widespread in the community. Immunocompetent people generally control the infection adequately and do not require medical intervention, though persistent latent infection is the norm[17]. In immunocompromised hosts, however, may have either reactivation of latent infection, or be unable to control a *de novo* infection. Toxoplasma encephalitis occurs when the CD4+ count falls below 100 cells/μl[17]. Pathological examination of toxoplasma lesions reveal a laminated structure consisting of three distinct zones[7]. First, there is a central necrotic avascular zone, containing few organisms.

Surrounding this is a region engorged with blood vessels with numerous active organisms (tachyzoites). This region is very vascular and undergoes an intense inflammatory reaction with thrombosis of small vessels and migration of many inflammatory cells into it. Peripheral to this is again a relatively avascular zone containing fewer active organisms and more dormant forms (bradyzoites)[7,20]. MR features of cerebral toxoplasmosis vary depending upon activity of the lesion, immune status of patient and administration of steroids. These lesions, usually multiple, are characteristically located at the basal ganglia (most common), the thalami, and at the gray-white matter junction, and rarely the brainstem[21]. There is usually mass effect with extensive surrounding edema. Though they are usually T2 hyperintense, the characteristic target sign consists of a T2 hypointense lesion surrounded by vasogenic edema. They are hypointense on T1W sequences and enhance in a ring or nodular fashion on Gadolinium administration, typically ring enhancing with an eccentric nodule. Enhancement decreases with therapy or decreasing CD4+ cell count[21]. Hemorrhage is rare[7]. DWI shows increased diffusivity. Spectroscopy reveals marked elevation of lipid and lactate peaks. All other brain metabolites are markedly diminished. This is particularly important since the main differential on imaging as well as clinically is primary CNS lymphoma.

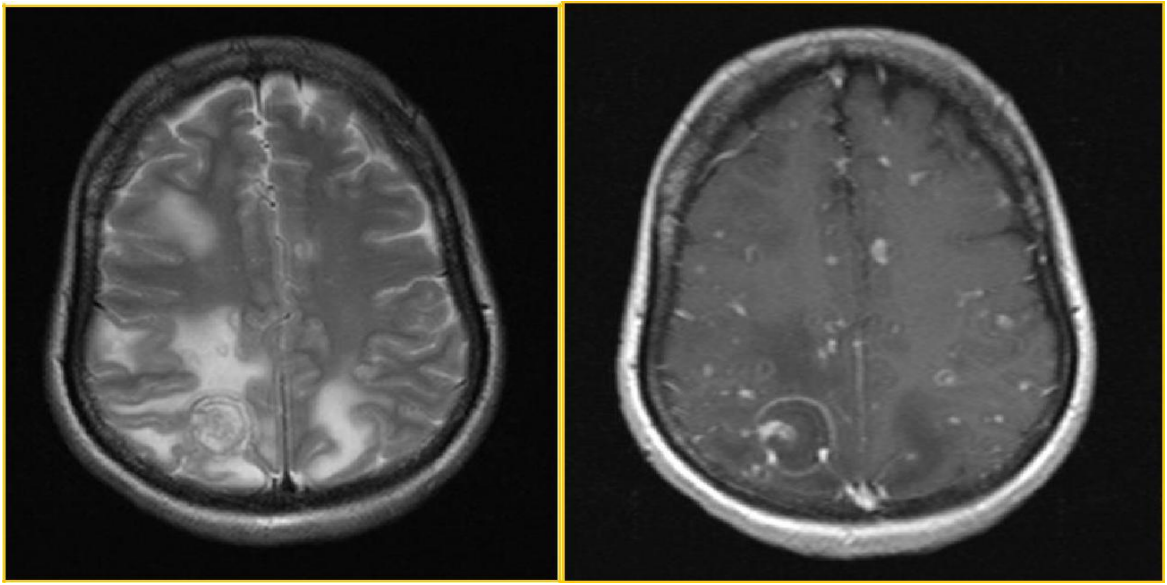

*Fig. shows multilple T2 hyperintense lesions throughout the brain, with the largest located in the right Parietal lobe with extensive surrounding edema, which shows peripheral enhancement. These are typical findings in Toxoplasmosis.*

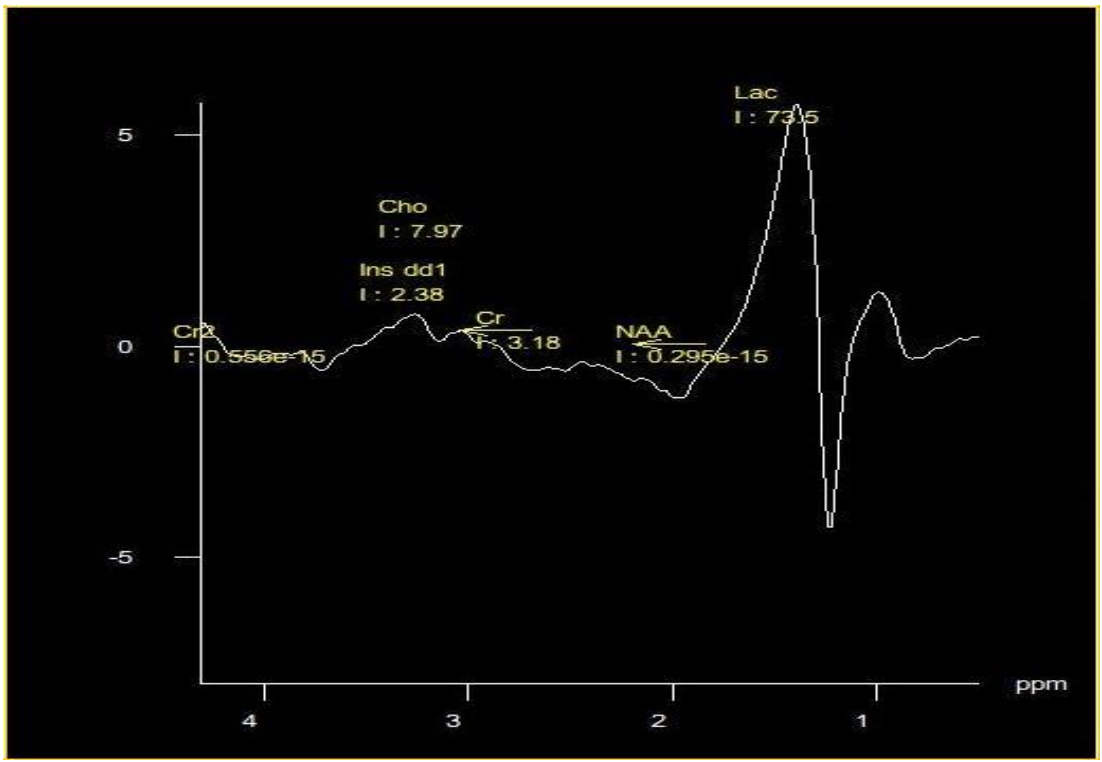

*Fig shows Spectrum obtained from a Toxoplasma lesion, with reduction in all metabolites, but with a prominent lactate resonance.*

**Primary CNS Lymphoma**

PCNSL accounts for 1%–5% of all brain tumors and 1% of all NHLs in the body[22]. It develops almost exclusively in immunosuppressed populations such as in HIV positive patients, Organ transplant recipients, or inherited immunodeficiency disorders[23]. It is estimated that upto 3% of AIDS patients develop PCNSL[23]. They generally present as focal intraaxial masses though subarachnoid space involvement is common in recurrent disease[7]. Upto a quarter of patients may have concurrent or prior involvement of the vitreous of the eye[23]. Primary intraocular lymphoma has also been described, though it is rare[22]. The location of the lesion is usually supratentorial (75 to 80%), involving the deep gray matter, periventricular regions and corpus callosum, usually in contact with the ependyma. AIDS related PCNSL is multifocal in 30-80% of cases[22], however involvement outside the CNS is a late and infrequent occurrence[20]. Most primary brain lymphomas are of B-cell origin and diffuse large-cell B-cell lymphomas are the most common histologic group. In the setting of immunosuppression, the cells in nearly all such tumors are latently infected by Epstein-Barr virus, and this is useful in that markers of Epstein-Barr viral infection can be used as an aid in diagnosis. Within the tumor malignant cells infiltrate the parenchyma of the brain and accumulate around blood vessels. Reticulin stains demonstrate that the infiltrating cells are separated from one another by silver-staining material; this pattern, referred to as ―hooping,‖ is characteristic of primary brain lymphoma[20]. On unenhanced T1-weighted MR imaging, lesions are typically described as hypo-or isointense and on T2-weighted MR imaging, iso- to hyperintense, though these tumors may often be hypointense to gray matter[22]. Most lesions show moderate-to-marked contrast enhancement on Gadolinium administration. Isolated white matter hyperintensity on T2W sequences, or no contrast enhancement on T1- weighted MR imaging has also been described in some rare cases of PCNSL[22]. Other features of PCNSL that have been described on morphological MRI (Though for immunocompetent patients) are ―notch sign,‖ an abnormally deep depression at the tumor margin, and ―open ring enhancement[24]. There could be mild or marked perilesional oedema, which however is less than that of metastases or glioblastoma[7,25]. Hemorrhage or internal calcifications within the tumor are a quite rare finding.

Because CNS lymphomas are highly cellular tumors, water diffusion is often restricted, making them appear hyperintense on DWI and hypointense on ADC maps[24]. PWI shows an elevated rCBV, which helps differentiate from toxoplasmosis[7].

In PCNSL, MR spectroscopy demonstrates elevated choline and lipid / lactate peaks combined with high Cho/Cr ratios[22]. NAA peak is reduced. Cho/NAA and lactate/Cr ratios are also increased when compared to normal gray matter[26]. These can also be seen in glioblastoma multiforme and metastases but may help in differentiating PCNSL from other lesions, particularly toxoplasmosis.

**PML**

Progressive multifocal leukoencephalopathy (PML) is a fulminating opportunistic infection of the brain that occurs in approximately 4% of AIDS patients; it is caused by the JC papovavirus. Primary infection with JC virus occurs in childhood and remains latent. In the setting of severely altered cellular immunity, this virus causes extensive myelin breakdown and white matter destruction, because of its targeting of the oligodendrocytes, the cell responsible for the production of myelin[21].

On pathologic examination of the brain, patches of irregular, ill-defined destruction of the white matter ranging in size from millimeters to extensive involvement of an entire lobe of the brain are seen. On microscopic examination the typical lesion consists of a patch of demyelination, most often in a subcortical location, in the center of which are scattered lipid laden macrophages and a reduced number of axons. At the edge of the lesion are greatly enlarged oligodendrocyte nuclei with glassy amphophilic viral inclusions which contain viral antigens by immunohistochemistry. Within the lesions, there may be bizarre giant astrocytes with one to several irregular, hyperchromatic nuclei mixed with more typical reactive astrocytes. Imaging studies show extensive, often multifocal lesions in the hemispheric or cerebellar white matter. In fact a posterior location, though not pathognomonic is considered supportive evidence for a diagnosis of PML[21]. These lesions start subcortically[21] (the site of highest blood flow), then extend into the deep white matter. They demonstrate rapid progression in size, resulting in confluence of the lesions. Extensive white matter high signal on T2W /FLAIR imaging with sparing of the cortical gray matter is the hallmark of this infection. The lesions are initially round or oval but rapidly progress to become confluent and large, leading to poor prognostic outcomes characteristic of this aggressive disorder. Lesions of PML can be distinguished from those of TE by their nonenhancing character[8]. Contrast enhancement, however, especially when faint and peripheral, does not exclude a PML diagnosis[21] and has been considered by Berger et al to be one of the factors predictive of prolonged survival in AIDS patients with PML[27], along with higher CD4 counts, possibly because of the patient's ability to mount a better response. The lesions may show a central area of hyperintensity on T2W MR Images, resembling necrosis. It has also been demonstrated that cortical atrophy and ventricular dilation are not prominent features of PML, although follow-up studies have suggested some atrophic progression along with increased hypointensity on T1- weighted images[28].

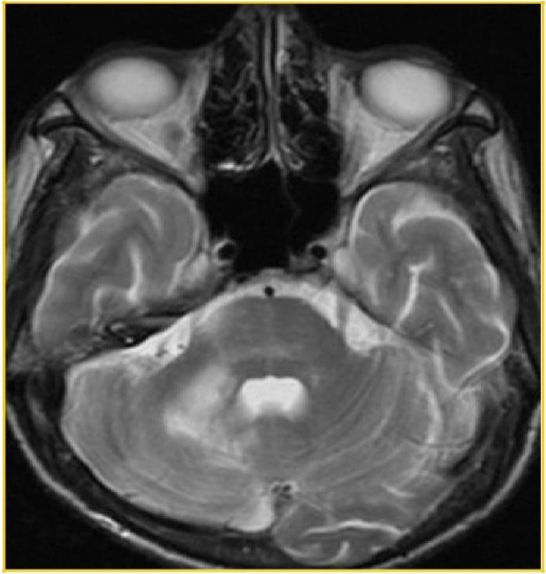
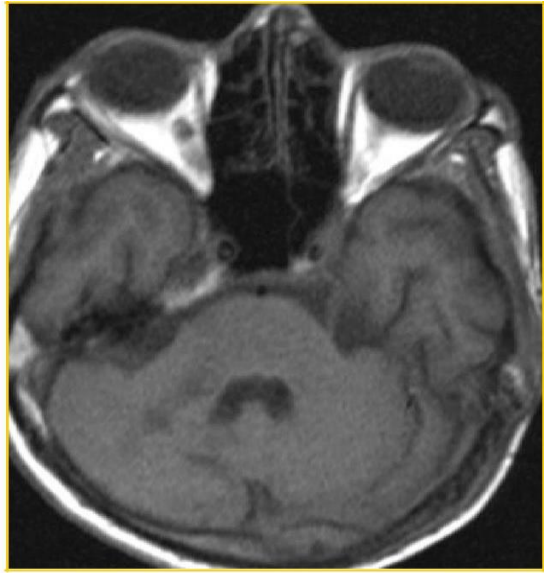
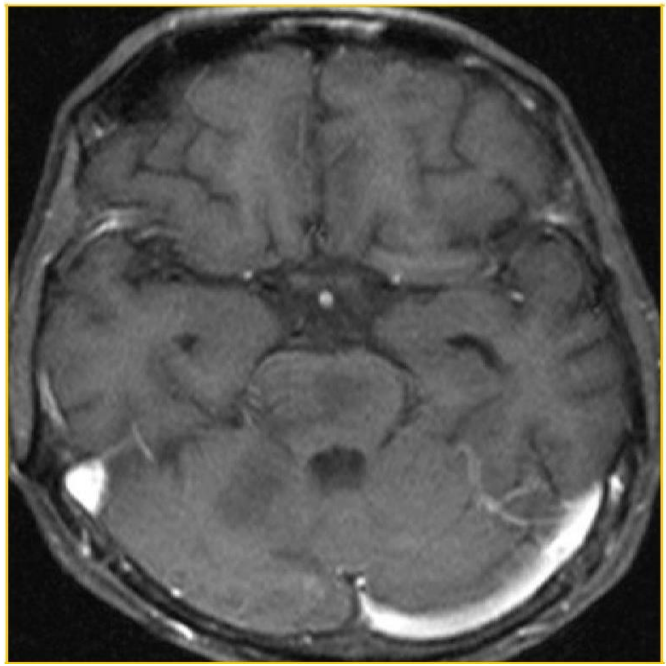

*Figures above show a T1 hypointense, T2 hyperintense lesion in the white matter of the right middle cerebellar peduncle. It shows no enhancement on post gadolinium images. This was a case of PML.*

Hyperintensity on FLAIR Sequence, the inversion recovery sequence which is used to null signal from fluid, is usually due to true progression of PML and is associated with a poor prognosis[28]. In contrast, hypointensity on FLAIR may represent a ―burnt-out‖ process of PML and is associated with longer survival.

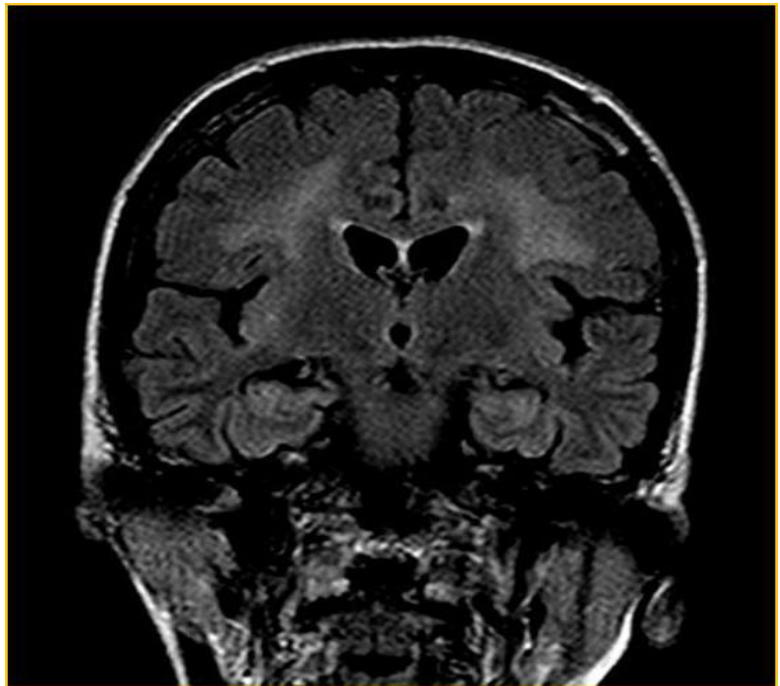

*Figure shows subcortical white matter hyperintensity in both parietal lobes sparing the cortical gray matter in a case of PML.*

PML is characterized on proton MR spectroscopy by a reduction in the NAA peak and elevation in the choline (Cho) and occasionally myoinositol (mI) peaks, with decrease in the NAA/Cr ratio and an increase in the lactate/Cr, Cho/Cr, and lipids/Cr ratios[29,30]. There are a few differences in the spectra obtained from patients with PML only, and PML with IRIS (Immune reconstitution inflammatory syndrome), but the most consistent in one large series is the mI/Cr ratio which was more elevated in patients with PML with IRIS[31]. The level of mIns depends upon the stage of the disease. In early and active stage, there is increased level of mIns that gradually decreases to normal level in late quiescent stage[32].

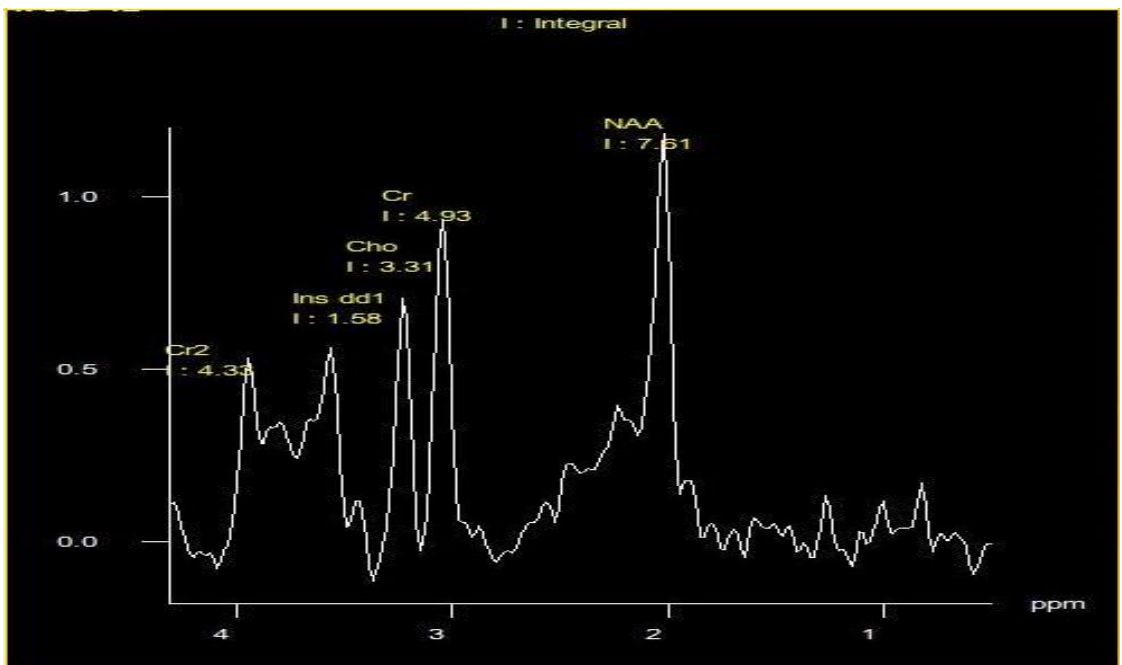

*Fig shows MR Spectroscopy obtained from a lesion in a case of PML reveals reduced NAA peak, along with prominent Choline and myoInositol peaks.*

# HIV Encephalopathy

Neurologic problems directly attributable to HIV occur throughout the course of infection and may be inflammatory, demyelinating, or degenerative in nature. The term *HIV- associated neurocognitive disorders* (HAND) is used to describe a spectrum of disorders that range from asymptomatic neurocognitive impairment (ANI) to minor neurocognitive disorder (MND) to clinically severe dementia[17]. The most severe form, *HIV-associated dementia* (HAD), also referred to as the *AIDS dementia complex*, or *HIV encephalopathy*, is considered an AIDS-defining illness. This consists of disabling cognitive, behavioral, and motor impairment associated with HIV and has been reported in up to 15% of HIV-infected adult patients[18], though approximately 50% of HIV-infected individuals can be shown to have mild to moderate neurocognitive impairment using sensitive neuropsychiatric testing[17]. While this is generally a late complication of HIV infection that progresses slowly over months, it can be seen in patients with CD4+ T cell counts >350 cells/μl. HIV-associated dementia is the initial AIDS-defining illness in only about 3% of patients with HIV infection and thus only rarely precedes clinical evidence of immunodeficiency.

The precise cause of HIV-associated dementia remains unclear, although the condition is thought to be a result of a combination of direct effects of HIV on the CNS and associated immune activation. HIV has been found in the brains of patients with HIV encephalopathy by Southern blot, in situ hybridization, PCR, and electron microscopy. Multinucleated giant cells, macrophages, and microglial cells appear to be the main cell types harboring virus in the CNS. Histologically, the major changes are seen in the subcortical areas of the brain and include pallor and gliosis, multinucleated giant cell encephalitis, and vacuolar myelopathy[17]. Less commonly, diffuse or focal spongiform changes occur in the white matter. Areas of the brain involved in motor, language, and judgment are said to be most severely affected. Rather than having a specific pathologic lesion as its correlate, this disorder is most tightly related to the extent of activated microglia in the brain, not all of which are necessarily HIV-infected[20].

Imaging in general has low sensitivity in identification of early disease due to the microscopic size of lesions. The most frequent radiological finding is atrophy of the brain[21,33]. On MRI, unilateral hyper intense lesions are usually present on T2W images, mostly in the frontal lobe, and progress to bilateral involvement. In the later stages, MRI reveals a diagnostic pattern of abnormalities: extensive and symmetric involvement of the deep white matter with typically spared cortical gray matter. Contrast enhancement and mass effect are not features of lesions in HIV encephalitis[33].

Among MRI techniques, magnetization transfer ratio (MTR) may be used to demonstrate macromolecular destruction (myelin and axonal loss) in HIV encephalitis and may help early diagnosis and in differentiation between PML by showing a degree of destruction which is more pronounced in PML[34].

MRS studies consistently reveal a severe reduction of NAA in the cortex and frontal white matter in those patients with HIV encephalopathy with more severe cognitive impairment. Furthermore, increasing levels of Cho and mI in the basal ganglia and the frontal white matter were also associated with an increasing severity of cognitive impairment and a greater decline in CD4 count[8]. With the onset of widespread HAART for HIV positive patients, studies began to focus on whether improvement of the metabolic parameters occurred. HAART was found to impove HIV encephalopathy in addition to systemic measures of HIV infection. MRS studies in these patients detected improvement of brain injury measured by cerebral metabolites, particularly the glial marker mI, in patients with early HIV encephalopathy after HAART[35,36]. In addition, the degree of improvement in clinical severity of the disease is related to the degree of recovery with mI[35].

Other studies were directed towards detecting onset of HIV associated neuronal damage in the brain. In a landmark study by Ratai et al[37], MRS was used to monitor metabolite abnormalities in the basal ganglia, frontal cortex, and white matter in SIV-infected macaques (An animal model of AIDS) from initial infection until 4 weeks post infection. At peak viremia 2 weeks postinfection, a significant decrease in NAA and NAA/Cr and a significant increase in Cr were found only in the frontal cortex, indicating a localized decline in neuronal integrity coupled with increased energy consumption in this region. Conversely, mI and mI/Cr levels increased across all three regions of interest, indicating microglial activation and astrogliosis and the resultant mounting of global inflammation regardless of the region of brain sampled.

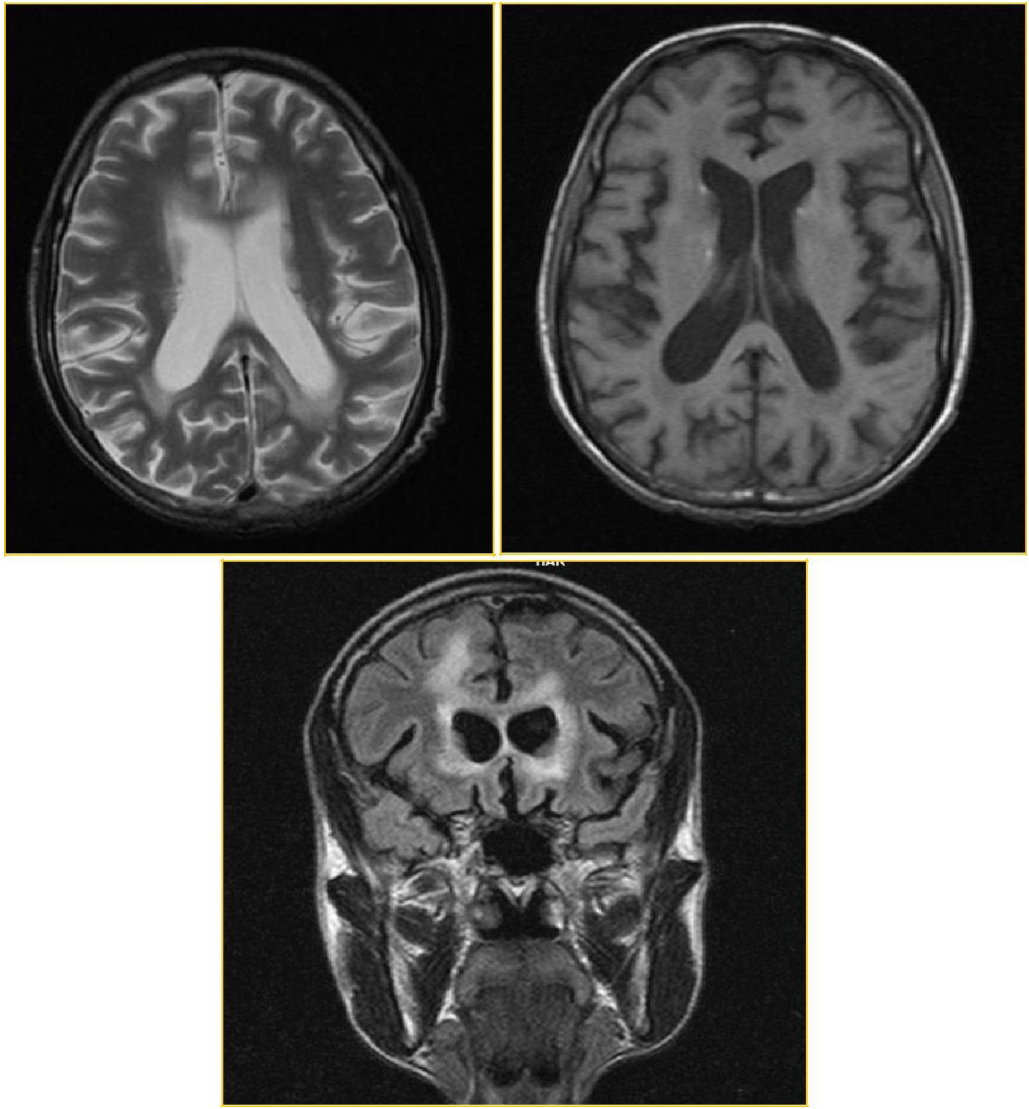

*Figures show deep white matter T2 and Flair hyperintensity with T1 hypointensity in both cerebral hemispheres in conjunction with prominent cortical sulci and ventricular system, findings seen in relatively advanced HIV Encephalopathy.*

Additionally, Cho and Cho/Cr levels increased initially in all three abovementioned regions at peak viremia, indicating macrophage infiltration, but then decreased below baseline by the fourth week. Further studies of patients with early HIV infection confirmed some of these findings including a sharp decline of NAA in the frontal cortex that was significantly correlated with expansion of CD8+ T cells.

Thus, neuronal injury occurs in the frontal cortex soon after infection with HIV[38], suggesting that an increase in NAA may serve as a surrogate marker for therapy effectiveness in early stage treatment regimens. Also, Patients with HIV associated neurocognitive disorders have also been found to have lower levels of Glx (Glutamine / Glutamate) concentrations and Glx/Cr ratio in FWM, which was associated with impaired performance in specific cognitive domains, including executive functioning, fine motor, attention and working memory performance[39].

MRS also shows alterations in the spectra of asymptomatic patients with HIV. In HIV patients, the NAA/Cho ratio was significantly lower, with lower values of NAA correlating with the CD4+ count[40]. A significant increase in Cho/Cr and mI/Cr ratios in asymptomatic HIV patients was also found, which increased further in patients with HIV encephalopathy[41]. The difference of NAA/Cr was more pronounced in the white matter than in the gray matter[42,43]. These results suggest that MRS is more sensitive than conventional MR imaging in detecting CNS involvement in neurologically asymptomatic HIV patients[42] and may, therefore, be used for early detection of brain damage induced by HIV.

Age and duration of HIV infection also, as expected, appear to have a significant impact on the brain. HIV and age interaction was associated with lower frontal white matter NAA. CVD risk factors were associated with NAA in both the gray and white matter in both groups. Past acute CVD events in the HIV positive patients were associated with increased mI in the cortex. Longer HIV duration was associated with lower NAA peaks in the grey matter.

While these findings are fundamental to our understanding of the underlying metabolic mechanisms behind the progression of cognitive decay in patients with HIV dementia, more research is needed to determine how HIV affects those in the early disease stage as well as asymptomatic subjects. MRS has proven to be a very useful tool in studies aimed at understanding how HIV infection affects the CNS.

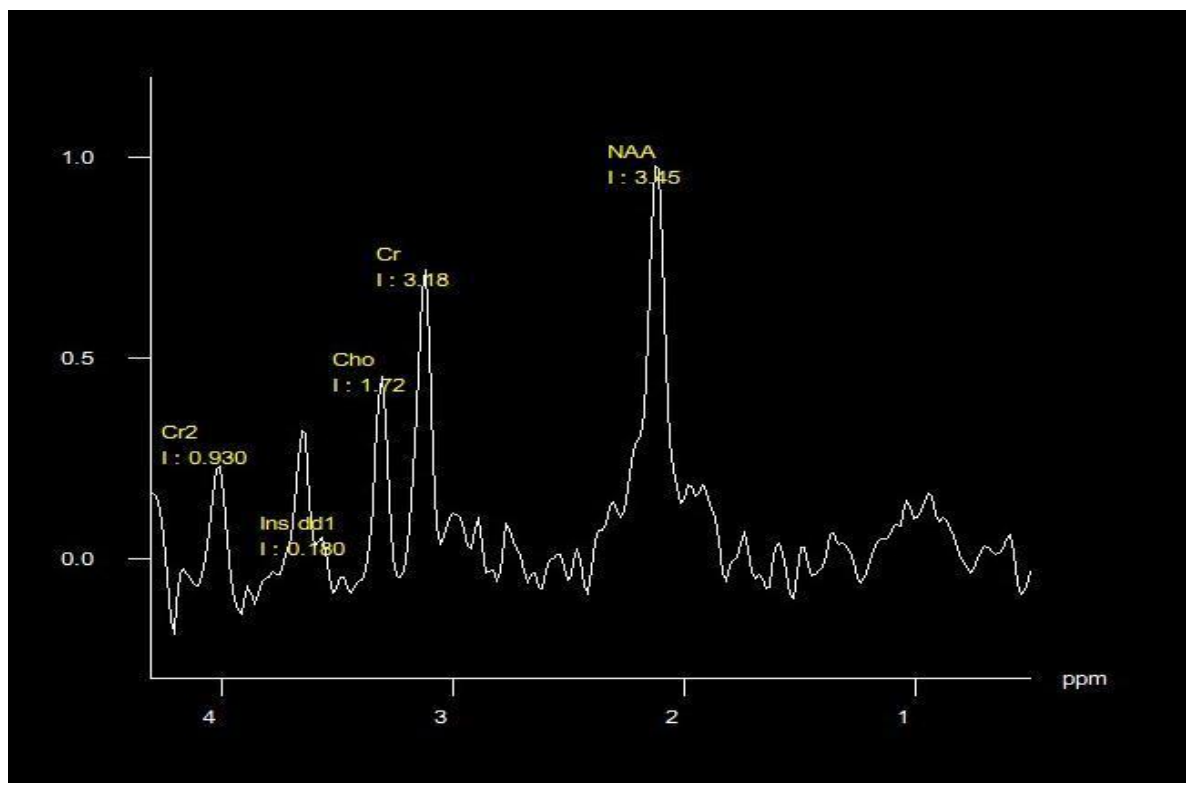

*Fig shows MR Spectrum obtained from the white matter in a case of HIV Encephalopathy, with reduced NAA peak and mildly raised choline peak.*

# Cryptococcosis

Cryptococcus neoformans is an opportunistic fungal pathogen that affects mainly immunocompromised patients. It is the most common fungus to involve the CNS and the most common fungal infection to occur in AIDS patients[7]. The primary site of infection is the lung, and infection is acquired via the inhalation of aerosolized spores[17] from soil or from pigeon droppings. The infection can present as meningoencephalitis, encephalitis, cryptococcoma or an abscess, of which meningitis is the most common manifestation[33].

Cryptococcus has yeast but not pseudohyphal or hyphal forms. The 5- to 10 μm cryptococcal yeast has a highly characteristic thick gelatinous capsule. Although the lung is the primary site of infection, pulmonary involvement is usually mild and asymptomatic, even while the fungus is spreading to the CNS. The major lesions caused by *C. neoformans* are in the CNS, involving the meninges, cortical gray matter, and basal nuclei. In immunosuppressed people, organisms may evoke virtually no inflammatory reaction, so gelatinous masses of fungi grow in the meninges or expand the perivascular Virchow-Robin spaces within the gray matter, producing the so-called soap-bubble lesions. In patients with protracted disease, the fungi induce a chronic granulomatous reaction composed of macrophages, lymphocytes, and foreign body type giant cells[20], causing mass lesion called cryptococcomas. Neuroimaging studies can reveal four basic patterns. First, parenchymal cryptococcomas may be seen. Secondly, there may be numerous clustered small, round or oval cystic foci that were hypointense or slightly hyperintense on T1W and hyperintense on T2W images and non-enhancing on postcontrast T1-weighted images, located relatively symmetrically in the basal ganglia bilaterally and in midbrain, representing dilated Virchow-Robin spaces filled with gelatinous exudate. These dilated VR spaces seen at the basal regions of the brain are also named as ―gelatinous pseudocysts‖ or ―soap bubble lesions‖, and they do not enhance with Gadolinium[44] as they do not disrupt the blood brain barrier and are rarely associated with inflammatory edema. Third, there may be multiple miliary parenchymal and leptomeningeal nodules. This is a more fulminant form of CNS cryptococcosis where there is disruption of the blood brain barrier and subsequent parenchymal seeding. These cryptococcomas appear as multiple enhancing leptomeningeal or parenchymal nodules. Fourth, a mixed pattern, consisting of dilated Virchow-Robin spaces with mixed lesions such as cryptococcoma and miliary nodules may be seen[45]. On diffusion-weighted images, there may be restricted diffusion in some of the lesions due to the high viscosity of their contents[44].

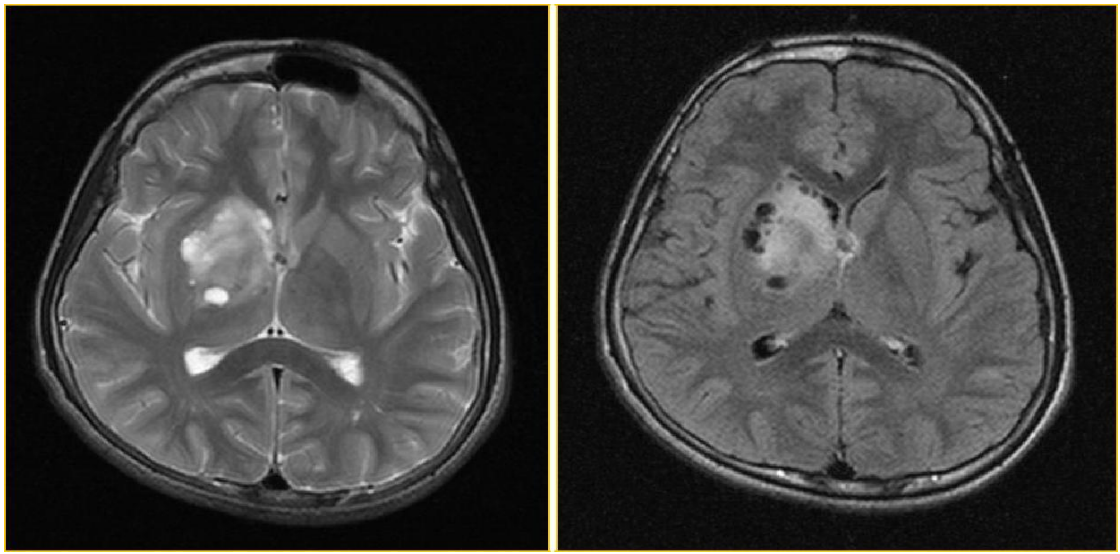

*Fig shows T2 hyperintense focal lesions in the right basal ganglia, which suppress on FLAIR sequence. These are dilated perivascular spaces seen in cryptococcosis.*

MR spectroscopy from cryptococcomas demonstrate a decrease in N-acetyl-aspartate (NAA), slight increase of choline (Cho), presence of lipids, trehalose and, in some cases, presence of lactic acid[44]. This correlates with an inflammatory lesion in the brain with neuronal destruction. Spectra of fungal abscesses in general are known to show peaks corresponding to lipids, lactate, alanine, acetate, succinate and choline[46]. Early in vitro studies showed that MR spectra of C neoformans and cerebral cryptococcomas--but not of other fungi, healthy brains, or gliomas--were dominated by resonances from the cytosolic disaccharide alpha, alpha trehalose MR spectra of *C. neoformans* and cerebral cryptococcomas--but not of other fungi, were dominated by resonances from the cytosolic disaccharide alpha, alpha trehalose[47] from the fungal capsule. Hence a remarkably high concentration of alpha, alpha trehalose in relation to other metabolites that are visible with MR spectroscopy is diagnostic of C neoformans[47].

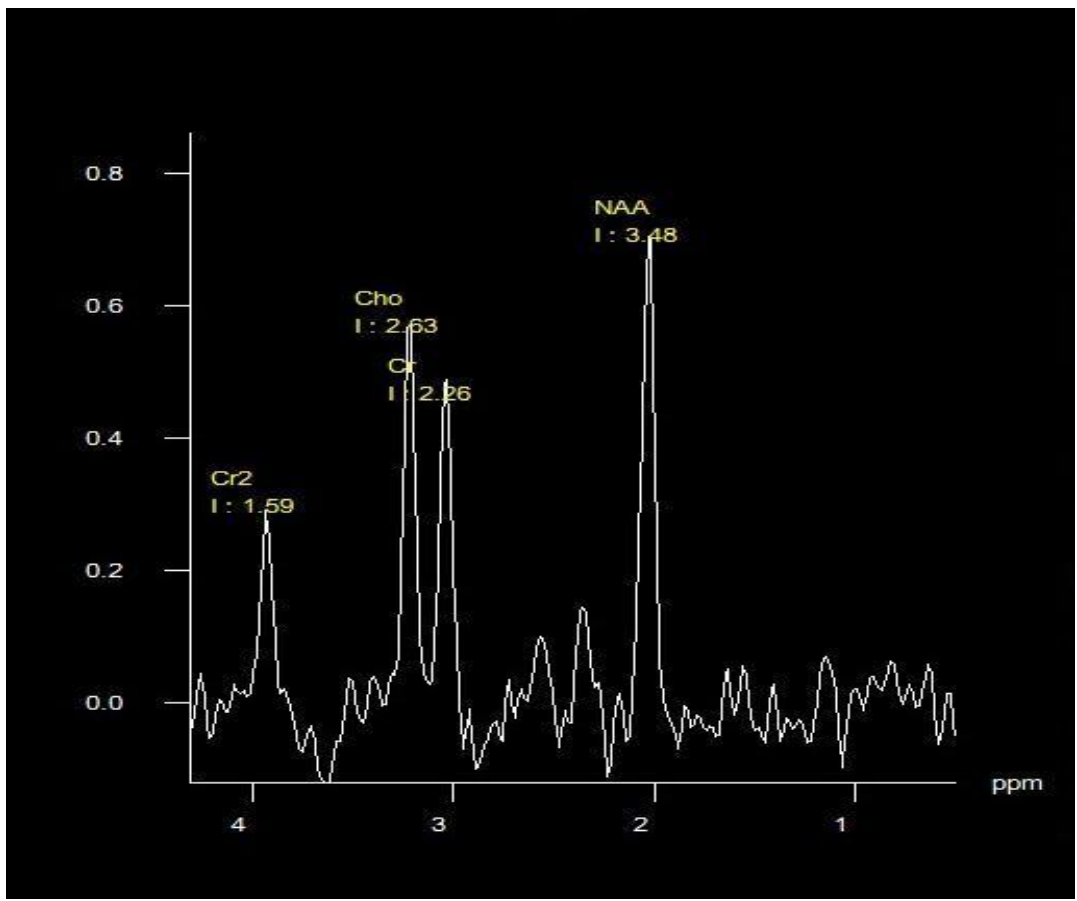

*Fig shows MR Spectrum obtained from a case of cryptococcosis with reduced NAA peak and raised choline peak, findings which are suggestive of ongoing inflammation..*

## CNS Tuberculosis

Central nervous system (CNS) disease caused by *Mycobacterium tuberculosis* is highly devastating, and accounts for approximately 1% of all cases of tuberculosis. It carries a high mortality and a distressing level of neurological morbidity, and disproportionately afflicts children and human immunodeficiency virus (HIV) infected.

The link between HIV and coinfection with tuberculosis runs deep. World Health Organization estimates that 9.27 million new cases of TB occurred in 2007 (139/100,000 population), compared with 9.24 million new cases (140/100,000 population) in 2006 world over[48]. India, China, Indonesia, Nigeria, and South Africa rank first to fifth in the total number of incident cases. Among the 15

Countries with the highest estimated TB incidence rates, 13 are in Africa, a phenomenon linked to the effect of high rates of HIV coinfection on the natural history of TB[49]. Incidence rates are falling world over except in Eastern Europe, where it is stable and increasing only in African countries with a low prevalence of HIV[48]. Among the 9.27 million incident cases of TB in 2007, an estimated 1.37 million (14.8%) are HIV-positive[48]. The global number of incident HIV-positive TB cases is estimated to have peaked in 2005 at 1.39 million. The relative risk of developing TB in HIV-positive people compared to HIV-negative people (the incidence rate ratio) is 20.6 in countries with a HIV prevalence greater than 1% in the general population, about 26.7 where HIV prevalence range from 0.1% and 1%, and 36.7 in countries with a prevalence less than 0.1%[50].

*Mycobacterium tuberculosis,* the bacillus responsible for tuberculosis reaches the CNS by the haematogenous route secondary to disease elsewhere in the body. CNS tuberculosis develops in two stages. Initially small tuberculous lesions (Rich's foci) develop in the CNS, either during the stage of bacteraemia of the primary tuberculous infection or shortly afterwards. These initial tuberculous lesions may be in the meninges, the subpial or subependymal surface of the brain or the spinal cord, and may remain dormant for years after initial infection. Later, rupture or growth of one or more of these small tuberculous lesions produces development of various types of CNS tuberculosis. The specific stimulus for rupture or growth of Rich's foci is not known, although immunological mechanisms are believed to play an important role. Rupture into the subarachnoid space or into the ventricular system results in meningitis. The type and extent of lesions that result from the discharge of tuberculous bacilli into the cerebrospinal fluid (CSF), depend upon the number and virulence of the bacilli, and the immune response of the host. Infrequently, infection spreads to the CNS from a site of tuberculous otitis or calvarial osteitis. The pathogenesis of localized brain lesions is also thought to involve haematogenous spread from a primary focus in the lung (which is visible on the chest radiograph in only 30% of cases)[51]. It has been suggested that with a sizeable inoculation or in the absence of an adequate cell-mediated immunity, the parenchymal cerebral tuberculous foci may develop into tuberculoma or tuberculous brain abscess.

Manifestations of CNS tuberculosis may be as follows[51]:

A. Intracranial

- Tuberculous meningitis (TBM), with or without miliary tubercles

- Tuberculous encephalopathy

- Tuberculous vasculitis

Space-occupying lesions: These might be tuberculomas (single or multiple), multiple small tuberculomas with miliary tuberculosis or tuberculous abscess

B. Spinal

- Pott's spine

- Tuberculous arachnoiditis (myeloradiculopathy)

- Intramedullary spinal tuberculoma

- Spinal meningitis

Tuberculous meningitis is not the most common but it is considered to be the most serious clinical form of extrapulmonary tuberculosis[52]. It is the most common cause of chronic meningitis, especially in developing countries[53]. In HIV-infected patients, TB is more likely to produce meningitis[54]. Common findings on imaging are abnormal meningeal enhancement in the basal cisterns, hydrocephalus, and vascular complications[53]. During the early stages of the disease, noncontrast MRI studies usually show little or no evidence of any meningeal abnormality. With disease progression, swelling of the affected subarachnoid spaces occurs with associated mild shortening of T1 and T2 relaxation times in comparison with normal CSF. Postcontrast T1W images show abnormal meningeal enhancement, especially in the basal cisterns. Cisternal enhancement is often quite striking and is far better demonstrated by postgadolinium MR than by CT[7]. Commonly involved sites are the interpeduncular fossa, pontine cistern, and the perimesencephalic and suprasellar cisterns. Involvement of the sulci over the convexities and of the Sylvian fissures can also be seen. Cerebellar meningeal and tentorial involvement is uncommon. The meningeal enhancement can also present as ependymitis. FLAIR sequence is more sensitive than T2 weighted images in depicting these abnormalities. Rarely, when the dura mater is involved, a focal thickening, known as pachymeningitis, associated with contrast enhancement develops and may or may not calcify. Hydrocephalus is a common sequela of tuberculous meningitis, and its presence is related to poor prognosis. The thick and gelatinous inflammatory exudate within the basal cisterns impedes the cerebrospinal fluid from circulating and being absorbed. Thus, communicating hydrocephalus develops, which can be documented by both CT and MR. The hydrocephalus occasionally may be obstructive, secondary to a focal parenchymal lesions with mass effect or due to entrapment of a ventricle by granulomatous ependymitis[7]. Periventricular hyperintensity on proton density and T2W images is due to the seepage of the CSF fluid across the white matter and usually suggests hydrocephalus under pressure, which is an indication for CSF diversion surgery to decompress the ventricular system. Chronic hydrocephalus may result in atrophy of the brain parenchyma. Endoscopic third ventriculostomy is a procedure gaining acceptance as patency of the stoma can be demonstrated by CSF flow dynamics[55] on MRI.

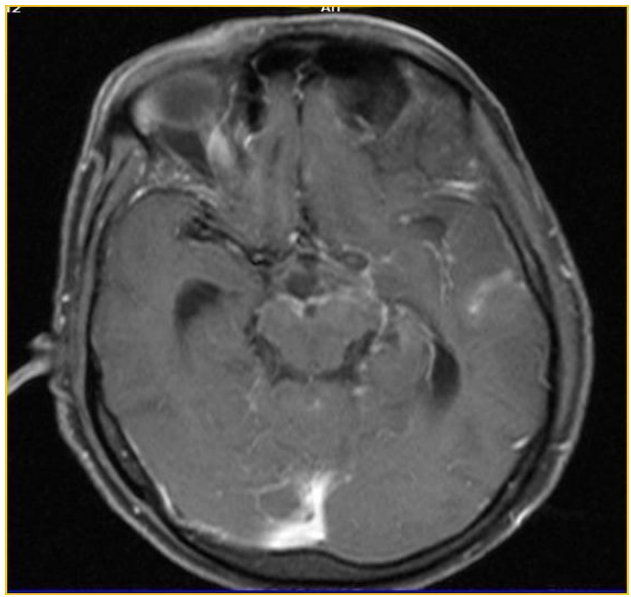

*Fig shows thick enhancing exudates in the interpeduncular and chiasmatic cisterns as well as the left sylvian fissure in a case of tuberculous meningitis.*

Miliary brain tuberculosis is usually associated with TBM. Miliary tubercles are <2 mm in size and are either not visible on conventional SE MRI images or are seen as tiny foci of hyperintensity on T2W acquisitions. After gadolinium administration, T1W images show numerous, round, small, homogeneous, enhancing lesions. Regular spin echo sequence invisible lesions that may or may not enhance after intravenous injection of gadolinium are clearly visible on MT-SE T1 weighted imaging. MT-SE imaging helps in defining the true disease load [53,56] in cases of miliary tuberculosis.

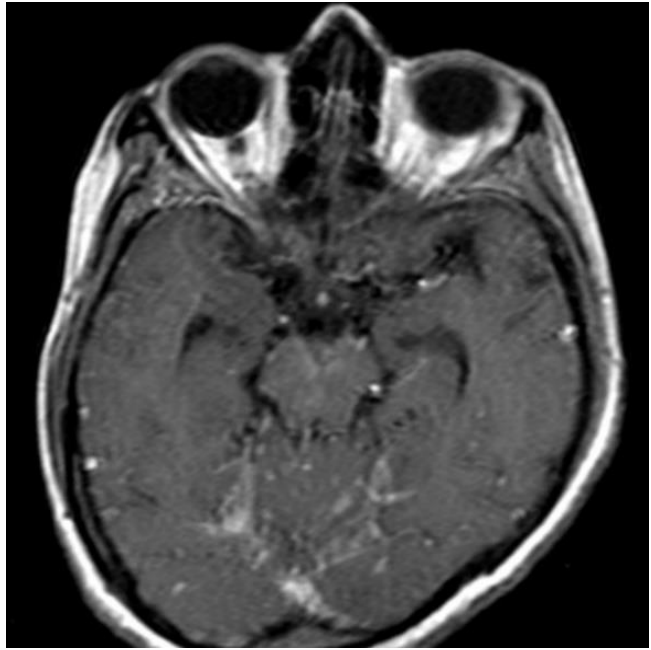

*Fig shows enhancing exudates in the supracerebellar cistern in this post gadolinium image with multiple tiny enhancing nodules throughout the brain parenchyma in a case of Tuberculous meningitis with miliary tuberculosis.*

Tuberculous encephalopathy is an established disease entity of diffuse cerebral damage occurring with tuberculosis with an underlying immune pathogenesis. Histopathology reveals demyelination and granuloma in the white matter. MRI shows diffuse, hyperintense lesions in the white matter on T2 weighted images, with gadolinium enhancement on T1-weighted images[57].

Vasculitis is a complication that is commonly seen at autopsy in cranial TBM. The adventitial layer of small and medium-sized vessels develops changes similar to those of the adjacent tuberculous exudates. The intima of the vessels may eventually be affected or eroded by fibrinoid-hyaline degeneration. In later stages, the lumen of the vessel may get completely occluded by reactive subendothelial cellular proliferation. Ischemic cerebral infarction resulting from the vascular occlusion is common sequelae of tuberculous arteritis[58]. The middle cerebral and lenticulostriate arteries are most commonly affected[58]. MRI angiography may help in the detection of vascular occlusion. High-field MRA with contrast is more sensitive than conventional MRA in the detection of occlusion of smaller vessels that are more commonly involved by the pathology. The majority of the infarcts are in the basal ganglia and internal capsule due to the involvement of the lenticulostriate arteries[58]. Haemorrhagic infarcts are common[7,58]. Diffusion-weighted imaging helps in the early detection of this complication.

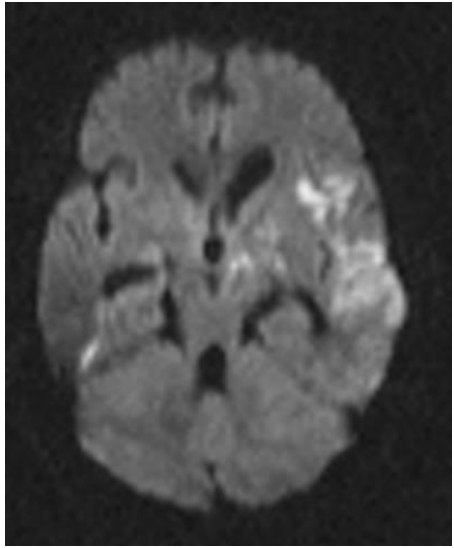

*Restricted diffusion is seen in the DWI sequence in the figure appended in a case of tuberculous meningitis. This is suggestive of vasculitic infarct.*

Magnetization transfer (MT) imaging is considered to be superior to conventional spin echo sequences for imaging abnormal meninges, which appear hyperintense on precontrast T1W MT images and show further enhancement on postcontrast T1W MT images[56]. In addition, MT ratio (MTR) quantification helps in predicting the etiology of the meningitis. Visibility of the inflamed meninges on precontrast T1W MT images with low MTR is specific for TBM, helping in differentiating it from other nontuberculous chronic meningeal infections[53].

There is no published study of *in vivo* MRI spectroscopy (MRS) in TBM; however, *ex vivo* spectroscopy of CSF has been attempted in this context[53]. High-resolution *ex vivo* MRS of the CSF shows signals from Lac, acetate, and sugars along with cyclopropyl and phenolic glycolipids. These have not been observed in pyogenic meningitis. The combination of *ex vivo* MRS with MT MRI may be of value in the diagnosis of TBM[53].

The most common parenchymal form of CNS TB is the tuberculoma[7]. Pathologically, a tuberculoma is composed of a central zone with caseous necrosis, surrounded by a capsule of collagenous tissue, epithelioid cells, multinucleated giant cells, and mononuclear in-flammatory cells[20]. It is mostly infratentorial when in children, and it is supratentorial in adults[53]. Tuberculomas may be single or multiple, and can be seen anywhere in the brain parenchyma. Tuberculomas are best viewed on T2 weighted images, and are classified based on their appearance in this sequence as[59]:

**T2 Hyperintense**

These lesions are noncaseating tuberculomas which are usually less than 1.5 cm in diameter. They appear T2 hyperintense and T1 iso to hypointense, hyperintense on MT T1 and FLAIR sequences. On contrast administration, they show nodular or ring enhancement. They are bright on DW images, with low ADC.

**T2 Hypointense**

These are lesions with conversion of the core into solid, cheesy, caseous material, and appear iso to hypointense on both T2 and T1 weighted images. These lesions may show an inconsistent non caseating peripheral region, which may be T2 hyperintense, but T1 iso to hyperintense, and this enhances on contrast administration. This peripheral region pathologically is found to consist of inflammatory cells, debris and mycolic acid and lipid rich cell wall fragments of *Mycobacterium tuberculosis*. This makes this peripheral rim hyperintense on MT T1 sequences. The central caseous material however has a High ADC, with no diffusion restriction seen on the DW images. Of importance is the presence of serine in the area of caseation on Spectroscopic analysis, which is a marker for the cell walls of *Mycobacterium tuberculosis* T2 hyperintense centre with T2 hypointense periphery: These lesions have a central core of liquefied caseous material. They may have a variably solid peripheral rim. This manifests on T2W images as a hyperintense lesion with a peripheral hypointense rim. The low signal rim in the external region is composed of collagen fibers produced by fibroblasts. Because the contents of collagen fibres vary, the low signal rim may be incomplete or absent[60]. T1W and MT T1 weighted images show this liquefied central area as hypointense, and DW images show diffusion retriction. There is rim enhancement on contrast administration.

Heterointense lesion: These lesions on histologic examination are relatively rich in inflammatory cells with scattered areas of cheesy caseous material. These show mixed intensity signal on T2 wighted images. On T1 weighted images also they are minimally hyperintense. MT T1 images show predominantly hypointense central regions with a hyperintense rim. On contrast administration, these lesions show variegated enhancement.

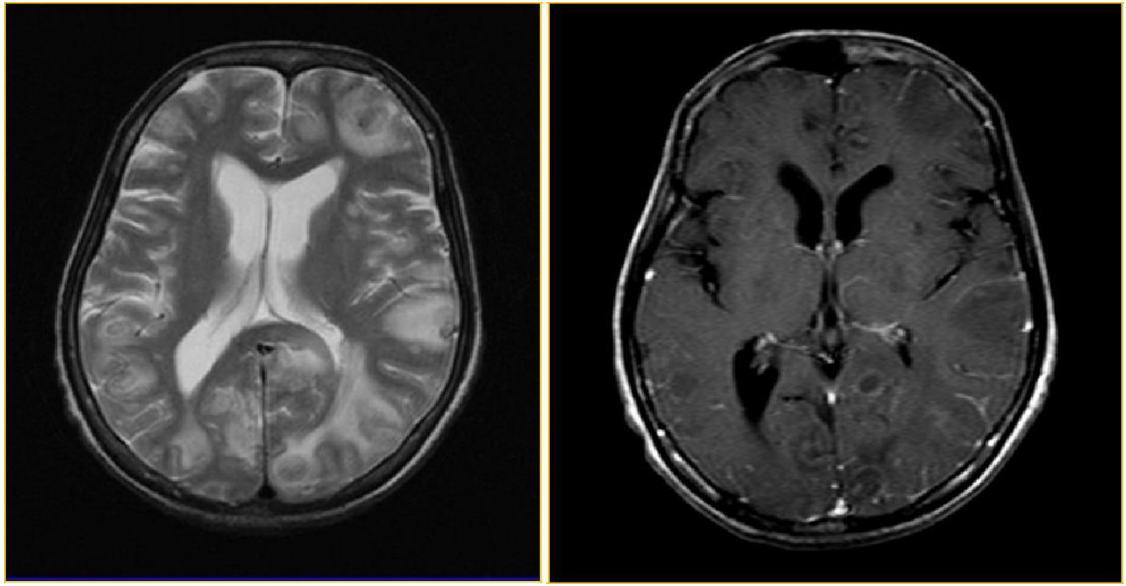

*Fig: The above figures show T2 hypointense ring enhancing (caseating) and T2 hyperintense disc enhancing lesions in an HIV positive patient s/o tuberculomas.*

In vivo spectra are found to be specific for intracranial tuberculomas and demonstrate the biochemical fingerprints of tubercle bacilli in a granuloma[53]. In vivo spectroscopy shows only lipid in T2 hypointense tuberculomas, whereas lesions with a heterogeneous appearance show Cho at 3.22 ppm along with lipid. These lesions show a large amount of cellularity and minimal solid caseation, the cellular regions appearing brighter on MT imaging and showing Cho resonance on spectroscopy. MR spectroscopy has shown prominent lipid and lactate peaks in patients with proven tuberculomas without aminoacid resonances. Caseous material typical of tuberculomas has a high lipid content. A pattern of multiple peaks for lipids have been documented corresponding to a terminal methyl group, methylene group, and phosphoserine. . Phenolic lipids represent the biochemical fingerprint of M. tuberculosis in a granuloma; however, phenolic glycolipids remain present in the virulent as well as non-virulent strains of M. tuberculosis. On ex vivo and in vitro (lipid extract) spectroscopy, peaks of cyclopropane rings and phenolic glycolipid caseating tuberculomas are seen, which can be attributed to M. tuberculosis[53]. This spectral pattern seems to be fairly specific for caseating tuberculomas because other lesions such as metastases and high-grade gliomas may show lipid peaks in addition to other metabolite peaks, whereas tuberculomas only demonstrate prominent lipid resonances.

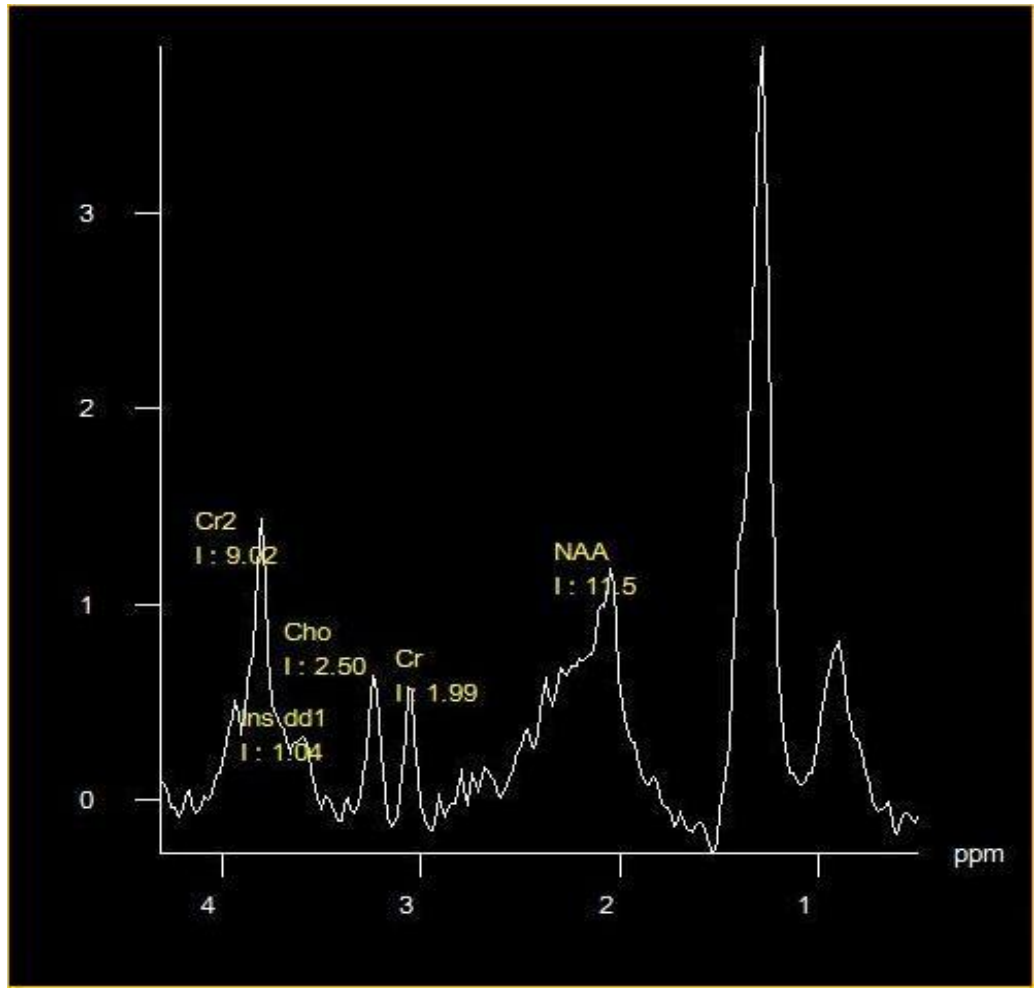

*Figure shows a spectrum obtained from a Tuberculoma which demonstrates great raised lipid-lactate resonance.*

Tuberculous brain abscess is a relatively rare condition caused by M. tuberculosis infection, constituting 4–7% of the total number of CNS TB cases in developing countries. According to the criteria of Whitener, tuberculous abscesses should show macroscopic evidence of abscess formation within the brain parenchyma and should offer histological confirmation that the abscess wall is composed of vascular granulation tissue containing both acute and chronic inß ammatory cells and M. tuberculosis[61].

Neuroimaging study findings are usually nonspecific for the detection of tuberculous abscesses. They appear as large, solitary, and frequently multiloculated, ring-enhancing lesions with surrounding edema and mass effect on MRI. MTR quantification from the rim of the abscess has helped in the diff erential diagnosis of tuberculous from pyogenic abscesses. High lipid-containing tuberculosis bacilli are probably responsible for the significantly lower MTR values in the rim of tuberculous abscesses compared with pyogenic abscesses. DWI in tuberculous abscesses shows restricted diffusion with low apparent diffusion coefficient (ADC) values, probably a result of the presence of intact inflammatory cells in the pus.

In vivo MRS has also been used for the differentiation of tuberculous abscesses from other lesions such as pyogenic abscesses and fungal lesions. In vivo proton spectra in tuberculous abscesses show only Lac and lipid signals without any evidence of cytosolic amino acids. More lipid peaks may also be apparent on ex vivo spectroscopy.

# Pyogenic abscess

Pyogenic brain abscess (BA) is defined as a focal infection within the brain parenchyma, which starts as a localized area of cerebritis, which is subsequently converted into a collection of pus within a well-vascularized capsule. Atypical brain abscesses are common in HIV infection[62]. Although infection with HIV-I does not predispose to pyogenic abscess formation, a considerable percentage of HIV-infected patients have acquired the infection and HIV patients, particularly those who are intravenous drug abusers are at risk for the development of pyogenic abscesses[33].

In a typical abscess with central liquefactive necrosis, the center of the cavity is slightly hyperintense to CSF, whereas the surrounding edematous brain is slightly hypointense to normal brain parenchyma on T1 weighted images. On T2 weighted images, signal intensities are quite variable. The rim is isointense to slightly hyperintense to white matter on T1 weighted images and is hypointense onT2 weighted images. The signal properties of the rim seen on MR have been attributed to collagen, hemorrhage, or paramagnetic free radicals within phagocytosing macrophages heterogeneously distributed in the periphery of the abscess. The sine qua non of abscess: smooth ring enhancement of an abscess capsule on post gadolinium MR images is the most important part of distinguishing a non-neoplastic from a neoplastic brain mass because nearly all enhancing brain lesions can present with ring enhancement. The abscess ring is most commonly very smooth, regular in thickness, and thin walled (approximately 5 mm in thickness), although it may be thinner along its medial margin, possibly due to variations in perfusion of gray and white mat-ter. Only infrequently, nodular or solid enhancement, incomplete thin rings, or thick and irregular rings may be observed, and these should provoke the consideration of other diagnoses. Daughter abscesses appear as adjacent smaller enhancing rings, often along the medial margin of the parent abscess, and can masquerade as a thick irregularity in the abscess wall at initial inspection. Abscesses tend to demonstrate high signal intensity on DWI, with a corresponding reduction in the apparent diffusion coefficient values. This is directly related to the cellularity and viscosity of the pus contained within an abscess cavity.

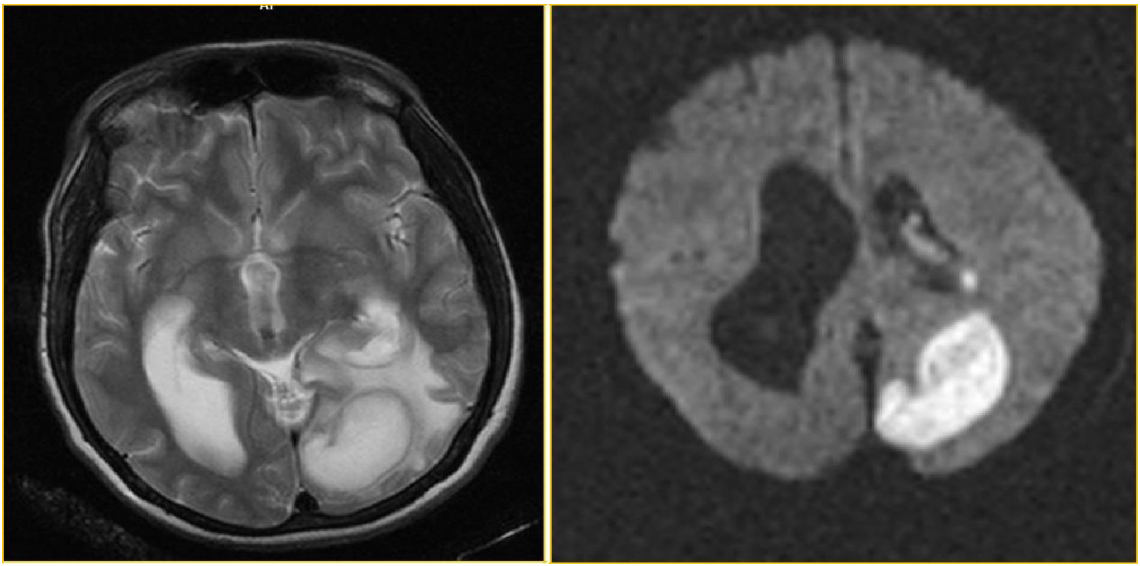

*Fig: shows a T2 hyperintense lesion in the left occipital lobe with a hypointense wall, showing restriction of diffusion on the DWI image, s/o brain abscess.*

The MR spectroscopy pattern derived from the central portion of an abscess appears to be fairly characteristic. Thus, MRS seems to contribute to distinguishing an abscess from a necrotic brain tumor and a tuberculoma. The presence of the resonance of cytosolicAAsis considered as a sensitive marker of pyogenic brain abscess[63]. In an untreated abscess, resonances may be seen corresponding to acetate, lactate, alanine, succinate, and pyruvate, as well as a complex peak indicating amino acids valine, leucine, and isoleucine[7]. Acetate, lactate, succinate, and pyruvate are metabolic end products arising from microorganisms. Acetate and succinate are not seen in association with necrotic tumors and are therefore fairly specific markers for pyogenic abscesses. However, these two resonances are not consistently identified in all abscess cavities. However, in vivo MRS does reveal an amino acid peak (valine, leucine, and isoleucine), which can be found in all abscesses but not in necrotic/cystic tumors. These amino acids arise from the proteolytic activity of neutrophils. In addition, the selective presence of Acetate with or without Succinate resonances on in vivo $^1$H-MR spectroscopy of abscesses has been described as a signature for anaerobic infection[63].

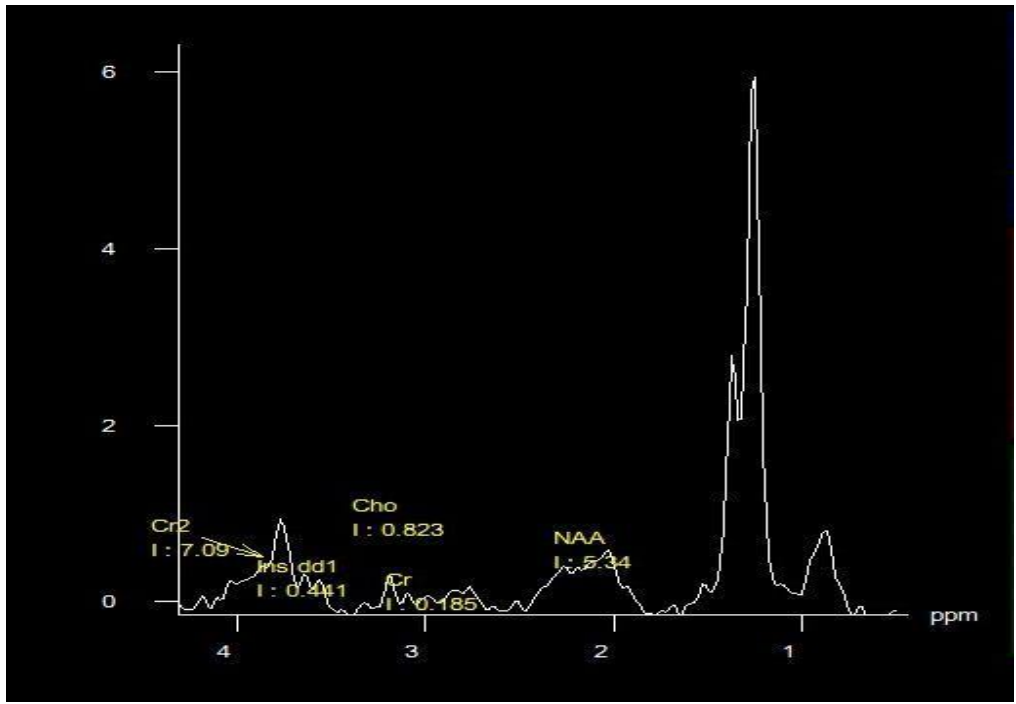

*Fig shows a spectrum obtained from a brain abscess revealing a giant lactate peak, complex resonance seen between 1.9 to 2.4 ppm represent cytosolic amino acids, confirming that this is a pyogenic brain abscess.*

## Cytomegalovirus (CMV) Encephalitis

In adults, CMV infection is typically seen in AIDS patients and is uncommon in other immunocompromised patients. CMV more often presents outside the CNS, involving the respiratory tract, liver, gastrointestinal tract, genitourinary tract, or hematopoietic system. This virus exists in a latent form in most of the population, and nearly 90% of adults have antibodies to CMV. Reactivation usually results in a sub-clinical or mild infection, mimicking mononucleosis. In some immunodeficient patients, however, reactivation results in disseminated infection and/or severe necrotizing meningoencephalitis and ependymitis. CMV may involve the CNS and/or peripheral nervous system[17].

MR imaging depicts patchy and, less often, confluent periventricular white matter lesions with hypointense signal on T1 weighted images, as well as a hyperintense signal onT2-weighted images. Infrequently, subependymal enhancement is evident and, if present, is a valuable diagnostic clue. Fat suppressed MR with gadolinium may reveal a thickened enhancing choroid and retina in patients with CMV retinitis[7].

MR Spectroscopy, while not pathognomonic, may show reduction in the N-acetylaspartate relative to creatine and an increase in the choline peak, and the presence of a double peak of lactate/lipid[7].

## Stroke

HIV patients have a small but definite increased incidence of stroke[64]. HIV infection can result in stroke via several mechanisms, including opportunistic infection, vasculopathy, cardioembolism, and coagulopathy[65]. HIV infection might predispose an individual to both arterial and venous thrombosis. However, the occurrence of stroke and HIV infection might often be coincidental. HIV associated vasculopathy describes various cerebrovascular changes, including stenosis and aneurysm formation, vasculitis, and accelerated atherosclerosis, and might be caused directly or indirectly by HIV infection, although the mechanisms are controversial. HIV and associated infections contribute to chronic inflammation. Highly active antiretroviral therapy is clearly beneficial in the management of patients with HIV infection, but impact the incidence of stroke in the HIV positive population in two important ways[65]. First, the drugs can be atherogenic and could increase stroke risk. Also, antiretroviral therapy can prolong life, increasing the size of the ageing population at risk of stroke.

The MR features of ischemic arterial stroke are well known, consisting, earliest, of hyperintensity on DWI with low ADC coefficient in a characteristic vascular territory. Subsequently developing features are cortical swelling and loss of gray white borders, with hypointensity on T1 weighted and hyperintensity on T2 weighted sequences. GRE (T2*) sequences are useful to detect foci of haemorrhage[10].

Venous sinus thrombosis manifests as loss of the normal flow void of the dural venous sinus. The thrombus in the sinus is initially isointense and hypointense on T1 weighted and T2 weighted sequences respectively and then grows to be hyperintense on both T1 and T2 weighted sequences. If a venous infarct occurs it appears hypointense on T1 and of mixed intensity on T2 weighted images, with mass effect due to extensive edema. Foci of haemorrhage are often present, readily detectable on GRE (T2*) sequences[10].

Metabolic abnorrmalities detected by MRS (reduced phosphocreatine and ATP, increased inorganic phosphate, elevated lactate, and acidosis) in ischemic brain due to arterial or venous thrombosis may precede structural alterations detected by any form of neuroimaging[66]. The characteristic feature is held to be Elevated lactate and decreased NAA[10]. An increase in tissue Lac production is seen with anaerobic metabolism and is detectable at a cerebral blood flow threshold just above that for tissue at risk of infarction[67]. Animal studies have suggested that Lac is detectable within minutes of stroke and that NAA may decrease by as much as 20% in the first hour and 50% in the first 6 hours. MRS may also help to refine penumbral definition by PWI–DWI mismatch. It has previously been hypothesized that preserved NAA but detectable Lac may represent the MRS penumbral signature. MRS may therefore be a reliable tool for the investigation of the pathophysiology of stroke[67].

## METHOD AND MATERIALS

**STUDY DESIGN** – Prospective observational study design.

The study was carried out for a period of 8 months in the Department of Radiology, Seth G.S. Medical College and K.E.M. Hospital, Parel, Mumbai with the approval of the Institutional Ethics Committee, IEC. Any HIV positive patients with neurologic disease referred for MRI examination on outpatients or inpatient basis was taken up for the study with the informed consent of the patient or guardian.
Those patients, who came to MRI for an unrelated pathology, were taken up as controls after taking consent from the patient.

## INCLUSION CRITERIA

- Patients who are HIV positive (by ELISA test) and are referred for MRI examination for neurologic disease.
- Patients able to cooperate for the examination e.g., ability to lie immobile for the period of the examination.

## EXCLUSION CRITERIA

Patient having absolute or relative contraindications for contrast enhanced MRI examination such as cardiac pacemaker, metallic (MRI non compatible) implants, or with kidney dysfunction (serum creatinine > 2mg/dl) etc.
- Patients who are unable to cooperate for the procedure.
- Patients or guardians who refused consent for the procedure.

Those patients who come to MRI for unrelated pathology were taken up as controls.

## STUDY PROCEDURE

36 consecutive HIV positive patients were examined by MRI and subsequently followed up for establishment of clinical diagnosis. The magnetic resonance imaging studies were conducted on Sonata Maestro class Magnetom 1.5 T MR Scanner from Siemens (Erlanger, Germany) with gradient field strength of $40mT/m^2$. A dedicated head coil was used. Unconscious patients, or patients with altered consciousness who came for the MRI scan were accompanied by a physician who pronounced the patients fit to undergo the scan, and was present during the study to monitor the patient. Occasionally sedation was required, administered by the accompanying physician.
Those patients who came for MRI for pathologies or symptoms unrelated to CNS, or with demonstrable normal scans were taken as controls. Proper informed consents were taken from the patients or their representatives as well as from the controls.

Standard protocol for brain MRI was -

- Acquiring the localizer.
- T2 tirm cor 3mm images.
- T2 TSE 3mm axial images.
- Gradient sequence
- T1 TSE sagittal 3mm images.
- Diffusion weighted imaging.
- Post Contrast T1 TSE coronal and axial 3mm images
- Multivoxel MR spectroscopy with short TE (TE=35ms)

Protocol was adjusted according to patient's requirement. Patients were followed up for final clinical diagnosis.
MR spectroscopy findings in any focal lesion detected in all cases were analysed and described ordering by the pathology the patients presented with and correlating with the final clinical diagnosis.

In all patients, the NAA/Cr ratio of the background white matter was recorded. Both cases and controls were distributed into two age groups: Patients below 40 years or equal to 40 years (Group 1) and patients greater than 40 years of age (Group 2). The NAA/Cr ratio was compared to that of the Age and Gender matched controls by means of statistical analysis.

**END POINT**

- To analyze the role of MR Spectroscopy as an important part of the MRI protocol in the diagnosis of HIV associated CNS disease.
- To compare the NAA/Cr ratio values of HIV positive patients and patients without HIV infection.

**OUTCOME MEASURES AND STATISTICAL ANALYSIS**

Analyzing the spectroscopy findings of patients, and comparing values of NAA/Cr ratio of cases with controls.

**RESULTS**

Plain and contrast MRI and MR Spectroscopy was done in 35 patients in this study. Patients from a wide range of age groups were included in the study and were divided into two groups: The first, with age less than or equal to 40 years comprised of 19 patients and the second with age more than 40 years comprised 16 patients. Amongst our patients, 24 were male and 11 were female.
Out of our 35 cases of CNS Disease in HIV positive patients, 8 patients had no identifiable focal or diffuse brain lesion. Out of the 27 patients who had identifiable lesions in the brain, 22 patients had infective disorders, 4 had vascular disease and 1 had neoplastic disease. The distribution of specific disorders was as follows: 6 had PML, 9 had CNS tuberculosis, 4 had toxoplasmosis, 2 had HIV encephalopathy, 2 had arterial territory infarcts, 2 had venous sinus thrombosis, 2 had cryptococcosis, 1 had meningioma, 1 had a brain abscess, and in 7 patients no focal lesions could be identified. One of these patients had both HIV Encephalopathy and Toxoplasmosis.

**Spectroscopy results –**
In controls, sequences of multivoxel spectroscopy with a TE of 30 ms were taken, at the level of centrum semiovale covering the subcortical white matter. In patients, sequences of multivoxel spectroscopy with a TE of 30 ms were taken covering the lesion and the adjacent white matter. Spectroscopic data were sampled from voxels within the lesions as well as from within the normal appearing white matter. The spectra from the white matter in cases and normal appearing white matter in controls were analyzed to obtain the NAA/Cr Ratio.

**DISCUSSION OF DATA**

The data at hand is continuous in nature. The sample size of NAA/Cr Ratio for Cases is 36 and that of NAA/Cr Ratio for Controls is 15. We have further divided the data in two categories, namely, (1) for people below (or equal) age 40 and (2) for people above age 40. All the tests which will be conducted here will have 95% level of significance.

**TEST FOR RANDOMNESS**

Any statistical analysis depends on the assumption that the data is necessarily random in nature. So we will now check for the randomness of the data by runs test (also called Wald–Wolfowitz test). This test will be conducted on NAA/Cr Ratio for Cases and NAA/Cr Ratio for Controls'. In both the tests our null hypothesis is that the data is random against the data is not random.
On applying the Runs test for NAA/Cr Ratio for Cases
p-value = 0.635
On applying the Runs test for NAA/Cr Ratio for Controls
p-value = 0.430
In both the cases as the p-values are greater than 0.05, we can conclude that the null hypothesis is true and hence the data is random.

**TEST FOR NORMALITY**

Now we have 3 pairs of data at hand. We will now test whether the data follow normal distribution (Gaussian). This test is very important as depending on the result, our test procedures will vary. We will use Anderson-Darling Test of Normality in all the 6 cases. 6 test for normality plots (Probability Plots) will be given below, each having the p-values given on those plots.

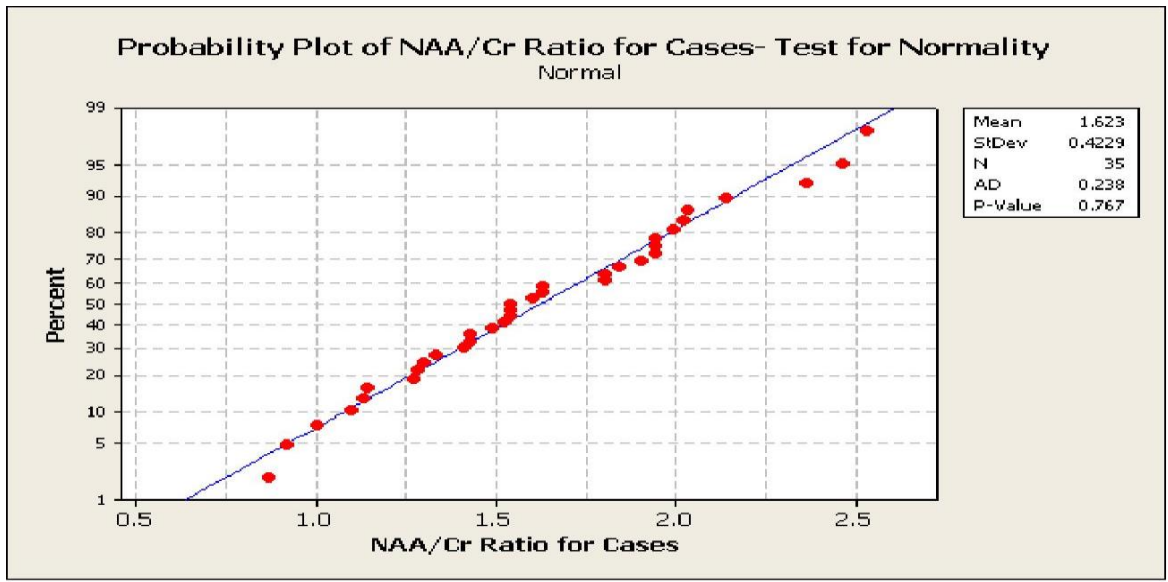

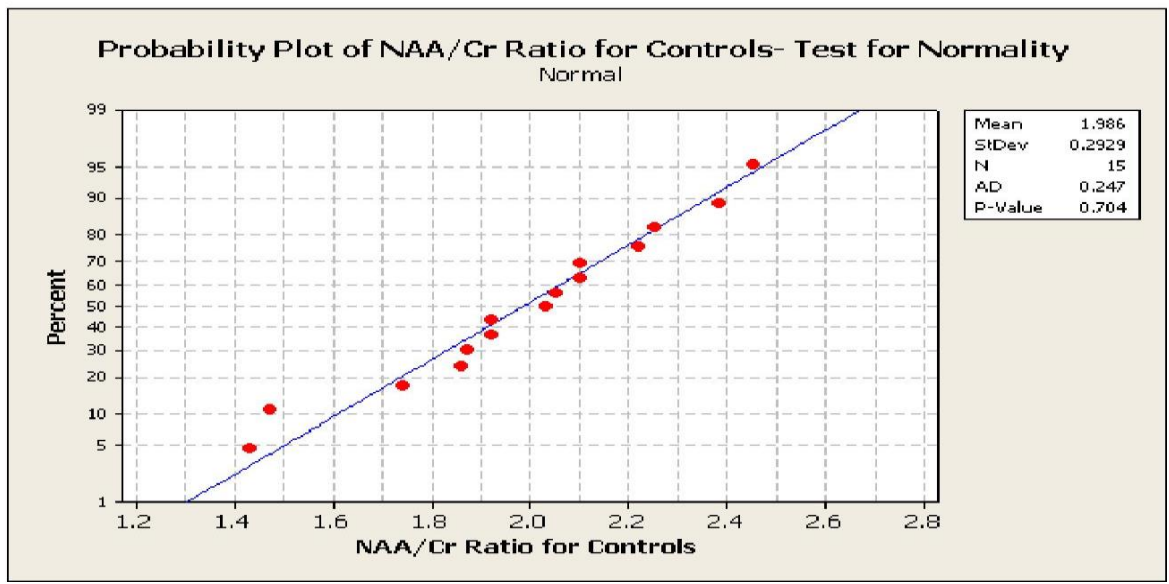

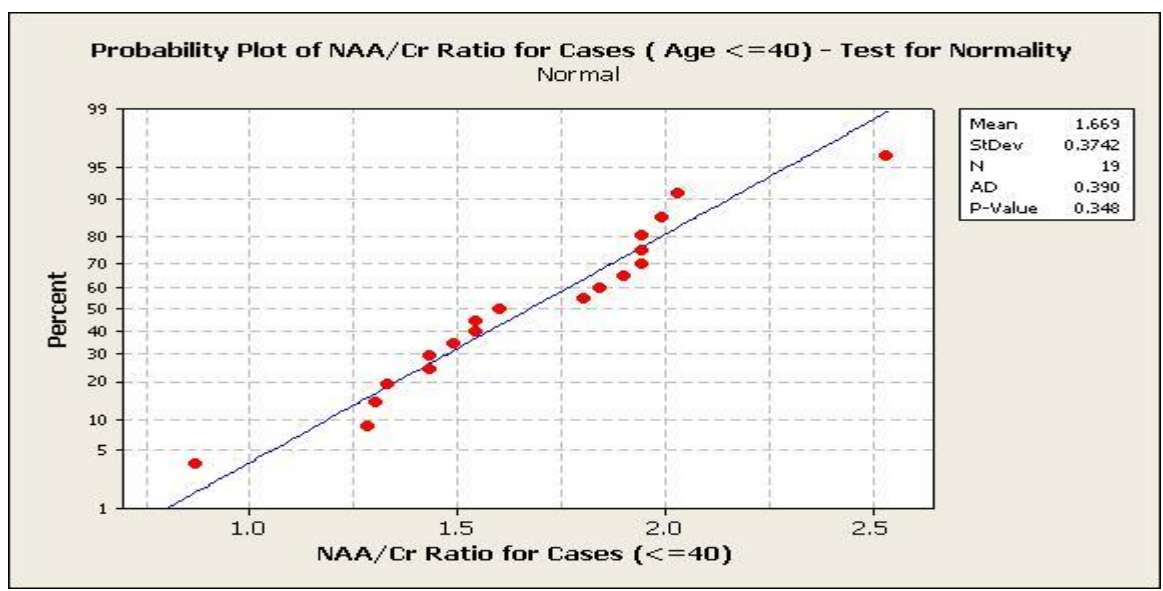

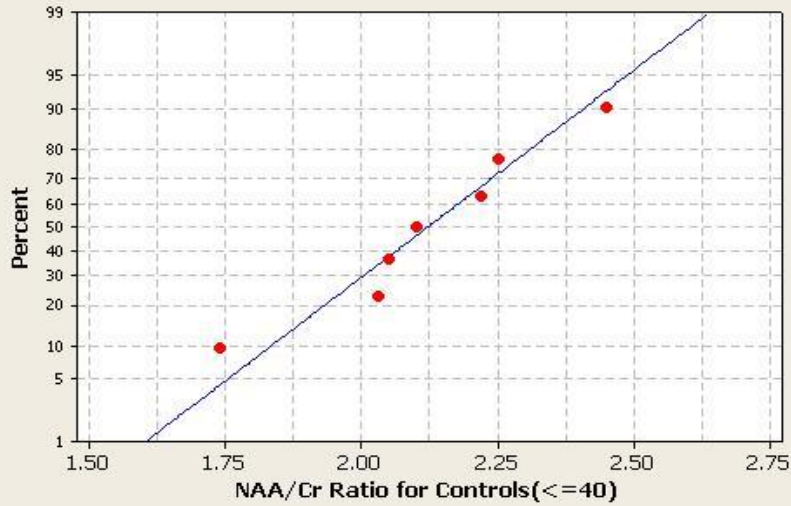
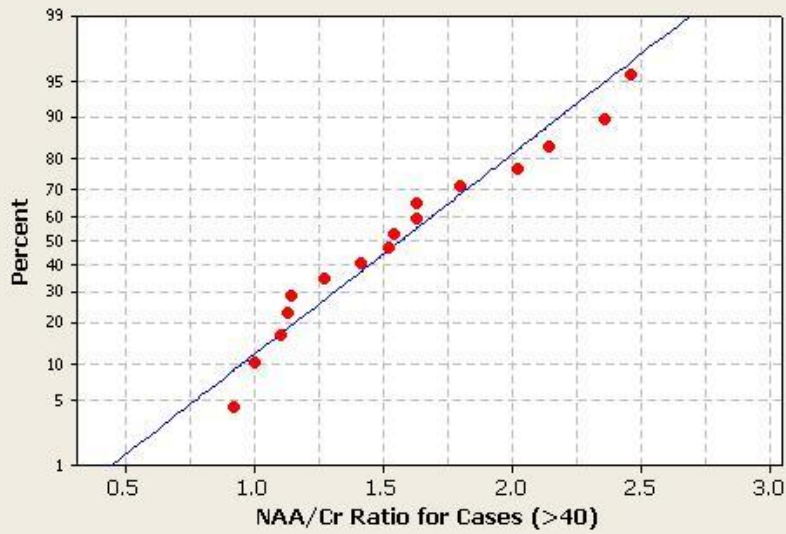
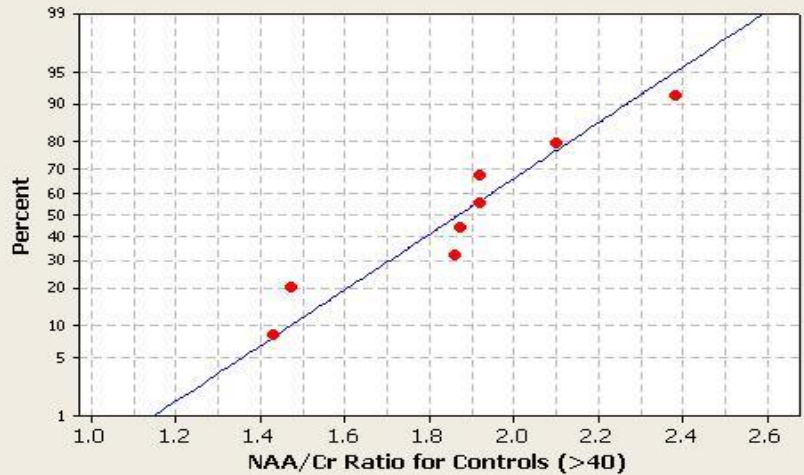

From the graphs it can be shown that all of them have p-value much greater than 0.05 and hence it can be concluded that the data follow Gaussian distribution.

**SUMMARY AND DESCRIPTIVE STATISTICS**

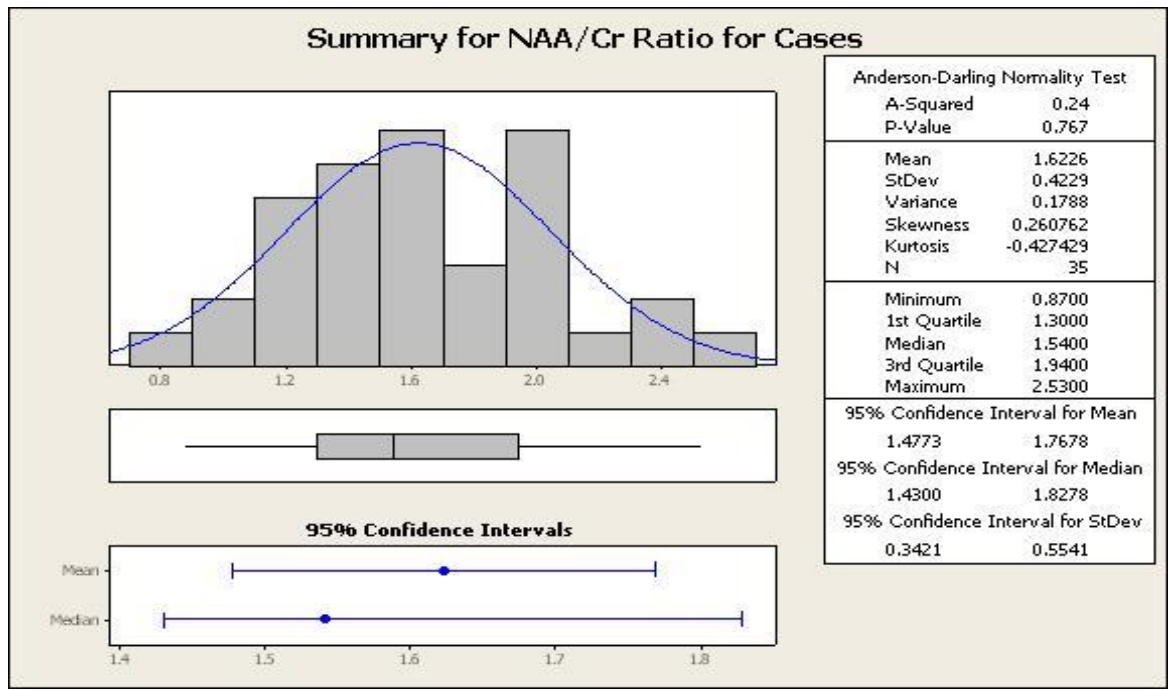

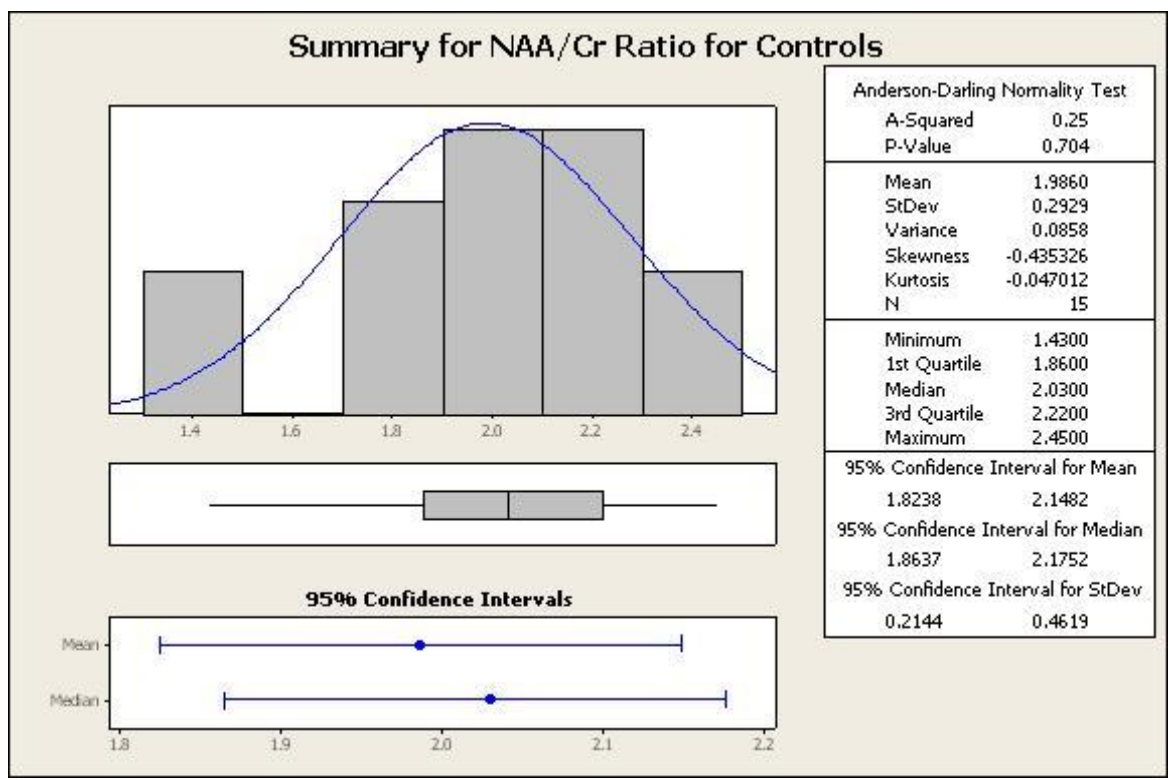

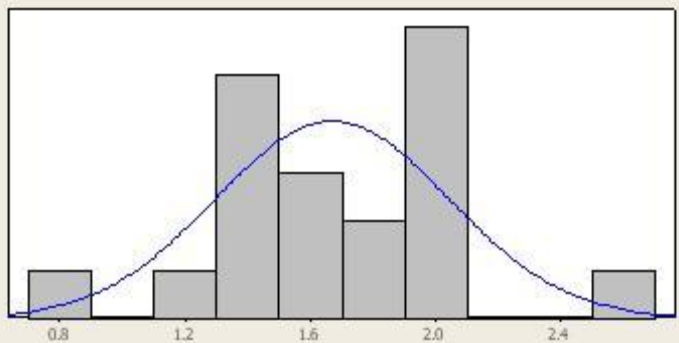
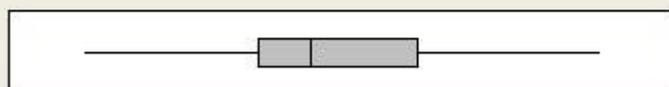
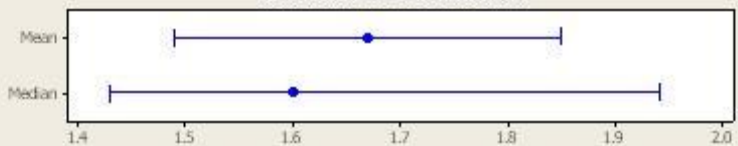

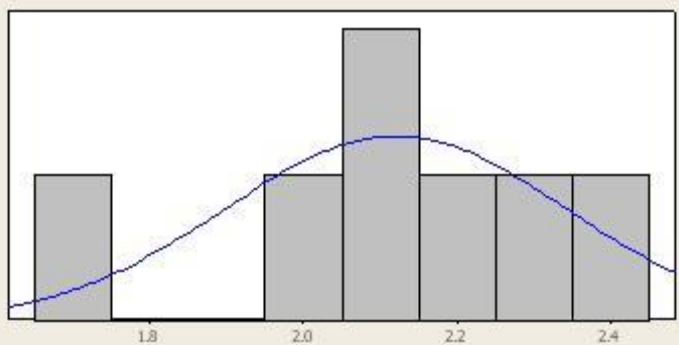
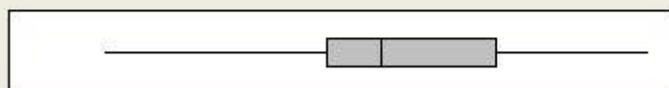
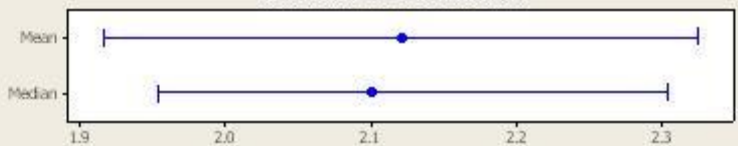

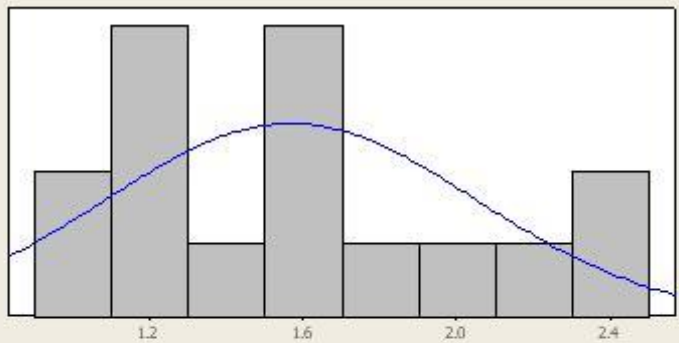
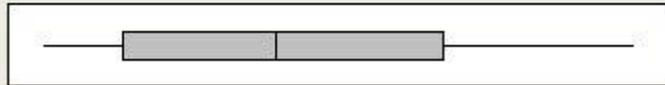
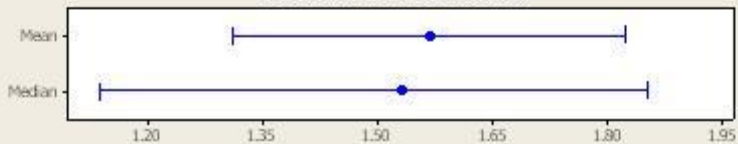

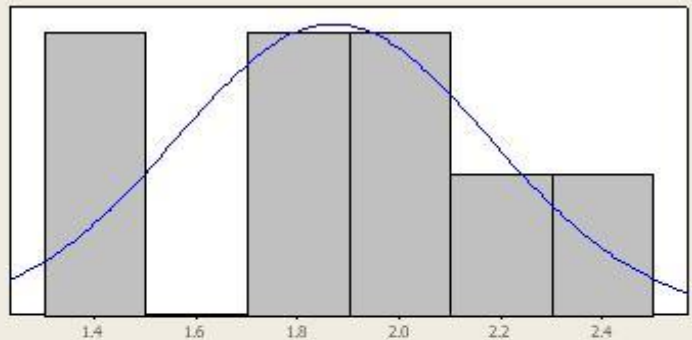
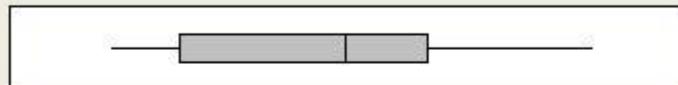
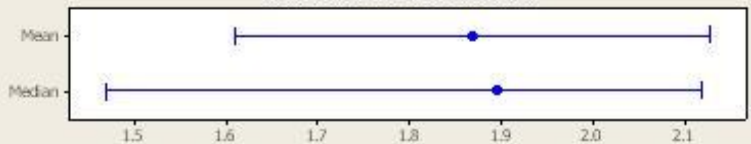

# TESTING OF HYPOTHESES

As we have 3 pairs of data, we will conduct three separate tests of mean of the ratio

- NAA/Cr Ratio for Cases vs. NAA/Cr Ratio for Controls
- NAA/Cr Ratio for Cases (<=40) vs. NAA/Cr Ratio for Controls(<=40)
- NAA/Cr Ratio for Cases (>40) vs. NAA/Cr Ratio for Controls (>40)

In all the cases, we will use _Two sample Welch's t-test as the data follows normality and the population variance is not known (which will be approximated by the sample variance in all the cases). We don't assume that the two populations under each test have homoscedasticity (or equal variance).

## 1. NAA/Cr Ratio for Cases vs. NAA/Cr Ratio for Controls

Our null hypothesis is mean of NAA/Cr Ratio for Cases and mean of NAA/Cr Ratio for Controls is equal against mean of NAA/Cr Ratio for Cases is lesser than mean of NAA/Cr Ratio for Controls

**Two-sample Welch's t test for NAA/Cr Ratio for Cases vs. NAA/Cr Ratio for Controls**

|  | N | Mean | St Dev | SE Mean |
|---|---|---|---|---|
| NAA/Cr Ratio for Cases (<=40) | 35 | 1.623 | 0.423 | 0.071 |
| NAA/Cr Ratio for Controls(<=40) | 15 | 1.986 | 0.293 | 0.076 |

t-Value = -3.49 (value of the statistic), p-Value = 0.001
Here the p-value is lesser than 0.05, so we reject the null hypothesis. Hence we conclude that NAA/Cr Ratio for Cases is significantly lesser than NAA/Cr Ratio for Controls. (Level of significance is set at 95%)

## 2. NAA/Cr Ratio for Cases (Age </=40) vs. NAA/Cr Ratio for Controls (Age </=40)

Our null hypothesis is mean of NAA/Cr Ratio for Cases (<=40) and mean of NAA/Cr Ratio for Controls (<=40) is equal against mean of NAA/Cr Ratio for Cases (<=40) is lesser than mean of NAA/Cr Ratio for Controls (<=40)

**Two-sample Welch's t test for NAA/Cr Ratio for Cases (Age </=40) vs. NAA/Cr Ratio for Controls (Age </=40)**

|  | N | Mean | St Dev | SE Mean |
|---|---|---|---|---|
| NAA/Cr Ratio for Cases (<=40) | 19 | 1.669 | 0.374 | 0.086 |
| NAA/Cr Ratio for Controls(<=40) | 7 | 2.120 | 0.221 | 0.084 |

t-Value = -3.76, p-Value = 0.001

Here the p-value is lesser than 0.05, so we reject the null hypothesis. Hence we conclude that NAA/Cr Ratio for Cases (<=40) is significantly lesser than NAA/Cr Ratio for Controls. (Level of significance is set at 95%).

### 3. NAA/Cr Ratio for Cases (Age >40) vs. NAA/Cr Ratio for Controls (Age >40)

Our null hypothesis is mean of NAA/Cr Ratio for Cases (>40) and mean of NAA/Cr Ratio for Controls (>40) is equal against mean of NAA/Cr Ratio for Cases (>40) is lesser than mean of NAA/Cr Ratio for Controls (>40)

**Two-sample Welch's t test for NAA/Cr Ratio for Cases (Age >40) vs. NAA/Cr Ratio for Controls (Age >40)**

|  | N | Mean | St Dev | SE Mean |
|---|---|---|---|---|
| NAA/Cr Ratio for Cases (<=40) | 16 | 1.567 | 0.481 | 0.12 |
| NAA/Cr Ratio for Controls(<=40) | 8 | 1.869 | 0.310 | 0.11 |

t-Value = -1.86, p-Value = 0.039

Here the p-value is lesser than 0.05, so we reject the null hypothesis. Hence we conclude that NAA/Cr Ratio for Cases (>40) is significantly lesser NAA/Cr Ratio for Controls (>40). (Level of significance is set at 95%).

### BOX-PLOT (BOX-WHISKER DIAGRAM)

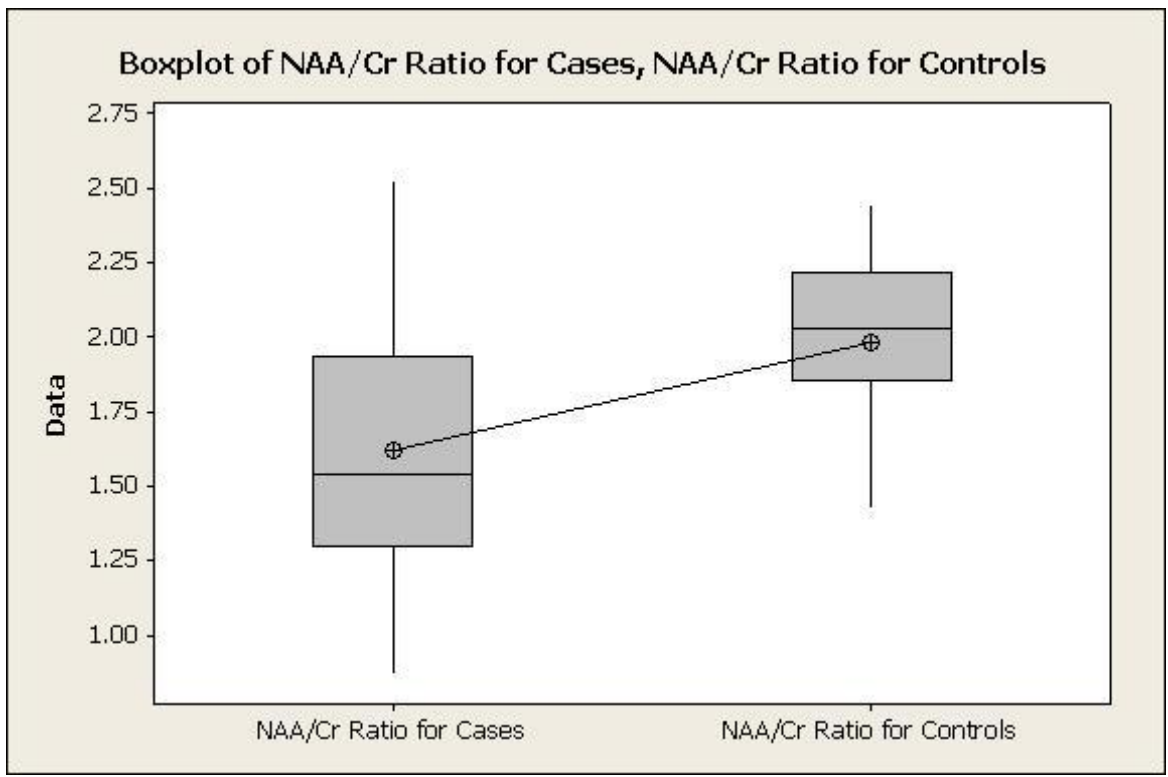

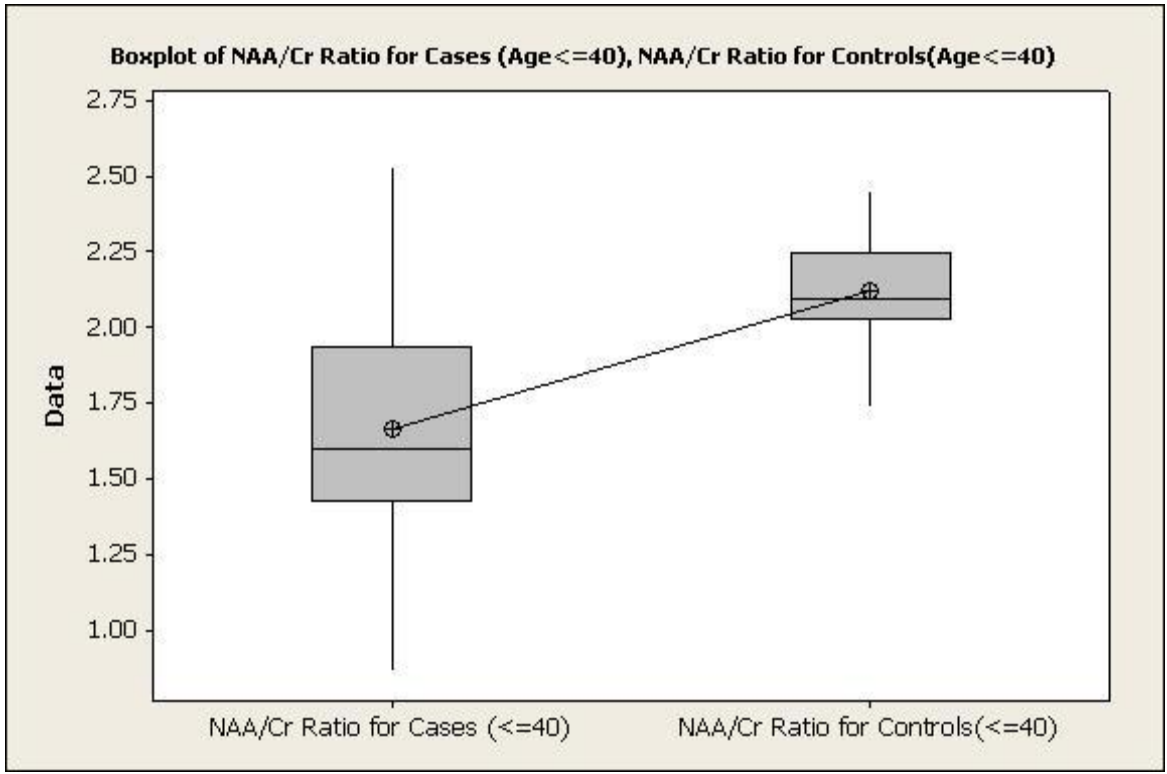

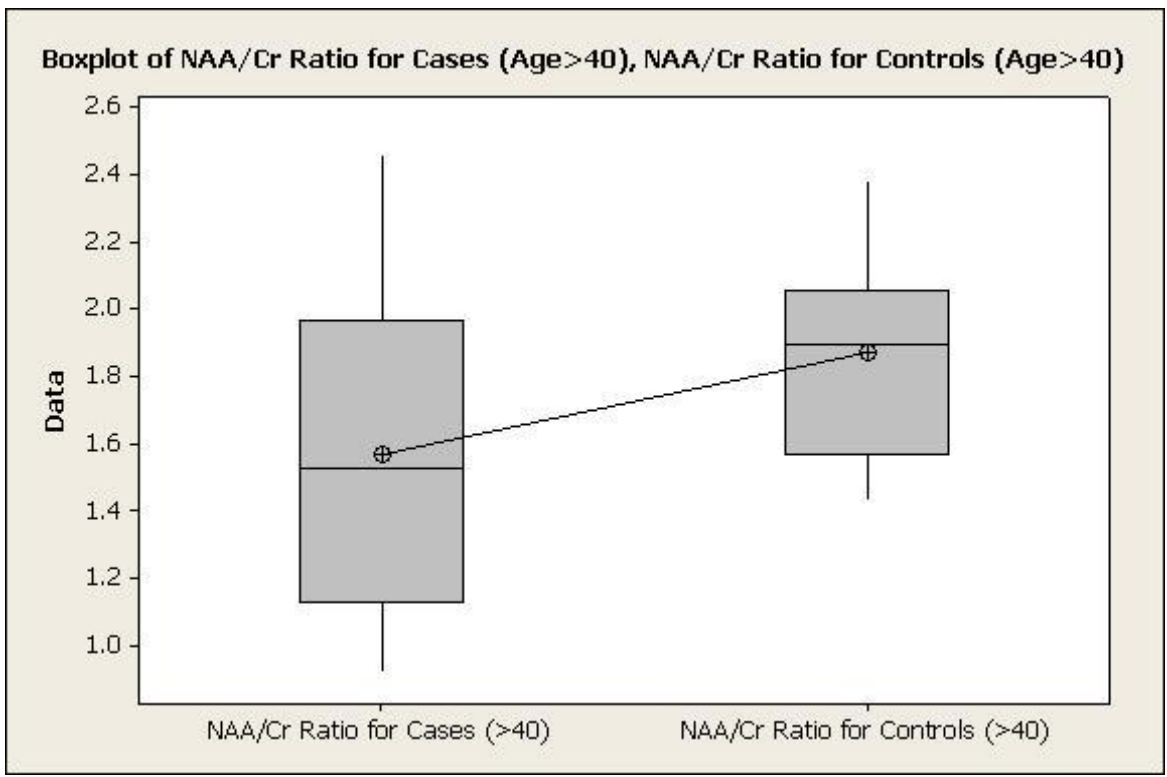

The specific parameters for PML were also tabulated.

| Number | Age (years) | Sex | Diagnosis | NAA/Cr Ratio | Cho/Cr | mI/Cr |
|---|---|---|---|---|---|---|
| 1 | 9 | M | PML | 1.94 | 0.7 | 0.24 |
| 2 | 40 | M | PML | 1.43 | 0.92 | 0.11 |
| 3 | 42 | F | PML | 1.1 | 0.88 | 0.1 |
| 4 | 44 | M | PML | 1.54 | 0.67 | 0.32 |
| 5 | 46 | M | PML | 1.63 | 0.9 | 0.16 |
| 6 | 55 | M | PML | 0.92 | 1.03 | 0.02 |

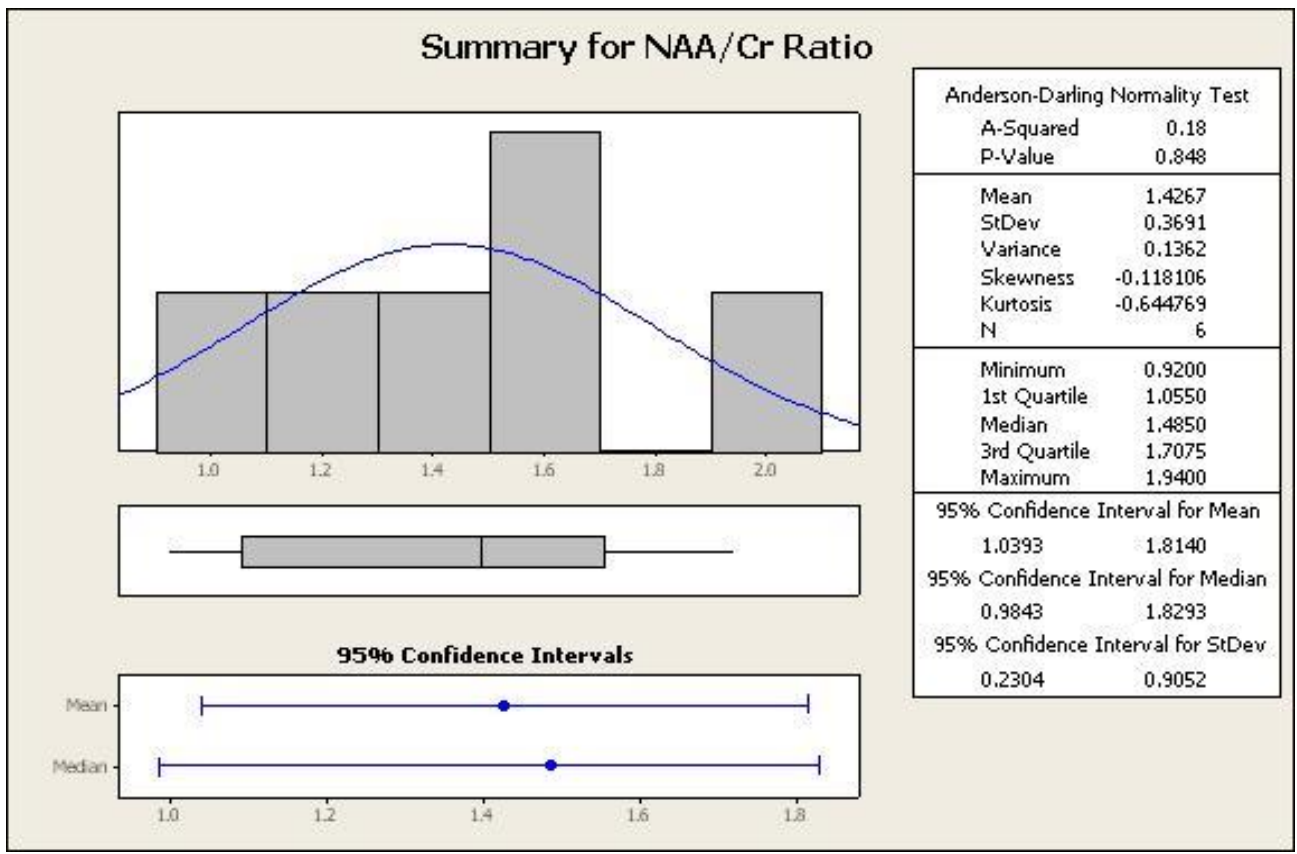

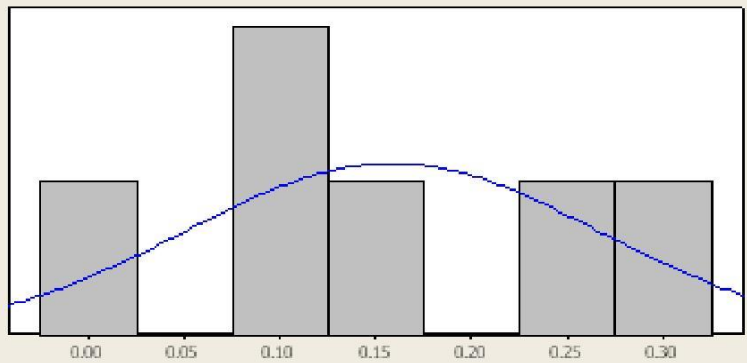
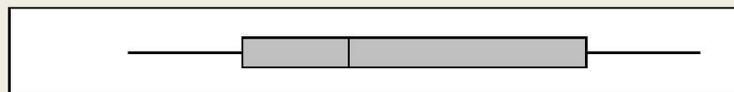
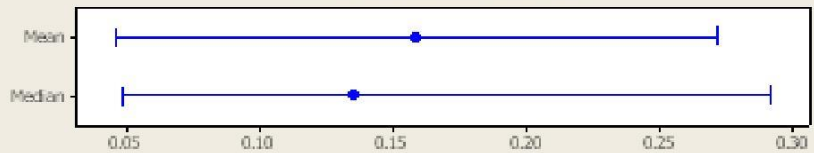

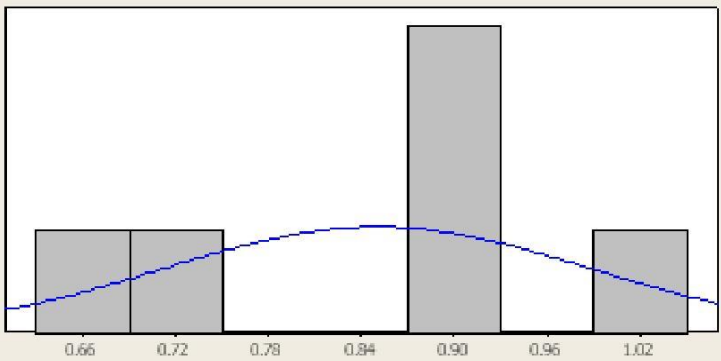
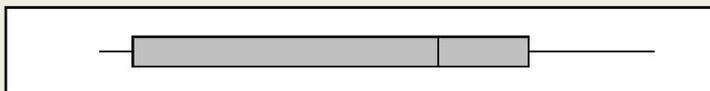
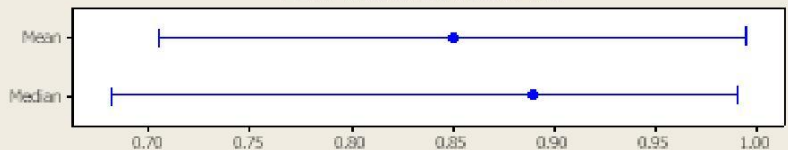

## DISCUSSION

MRI is undoubtedly the finest imaging modality for the evaluation of CNS disease in HIV and MR Spectroscopy has an important ancillary role in the establishment of the diagnosis. There have been multiple studies involving the distribution of CNS disease in HIV. HIV encephalopathy (AIDS Dementia complex) is commonly held to be the most frequent neurologic disease in the HIV positive population[7]. However unequivocal imaging evidence of HIV encephalopathy was found in only 2 of our patients (22%). **Collazos** found the most frequent CNS infections to be Toxoplasmosis, PML, Cryptococcosis and CMV infection[68]. A large study in Africa by **Luma et al** found Toxoplasmosis followed by cryptococcal infection to be the most common, and tuberculosis was also common[69]. In the East, in a large series in Thailand, **Kongsiriwattanakul** et al found that Cryptococcal infection was the most common infection, followed by tuberculosis[70], and in Korea **Choe et al** found PML and Tuberculosis to be the most common[71]. A recent Indian autopsy study by **Lanjewar et al** found Toxplasmosis to be the most common, with CNS Tuberculosis being almost as commonly found[72]. In our study, 25 patients (81%) were found to have neuroinfection, with the commonest being CNS Tuberculosis (9 patients, 24%), followed by PML (6 patients, 16%). Toxoplasmosis was the third most common seen in 4 patients (11%). Neuroinfection with tuberculosis takes various forms. In our study, 8 out of 9 (89%) patients were referred for MRI for tuberculous meningitis and one asymptomatic patient for follow up in a case of tuberculoma. 2 of the patients had vascultic infarcts (22%) and active tuberculomas were demonstrated in 2 (22%). No spinal tuberculous lesion was found in our study. MR spectroscopy is not of great significance in tuberculous meningitis; however spectra obtained from the tuberculomas revealed reduced NAA with a prominent lipid – lactate peak.
PML was the second most common pathology found in our study. PML is characterised spectroscopically by reduced NAA, increased choline and increased myoinositol. The NAA/Cr ratio is reduced in PML further than that due to the decrease caused by other pathologies in HIV positive population[73]. This was consistent with the values obtained in our study, where the mean NAA/Cr ratio for all HIV patients in our study was found to be 1.6226 with a standard deviation of 0.4229 and in cases of patients with PML, the NAA/Cr ratio 1.4267 with a standard deviation of 0.3691. The raised choline peak was also found in the spectra of all our patients (Mean Cho/Cr value 0.85).
Toxoplasmosis was the third most common neuroinfection in our study. While toxoplasmosis has characteristic imaging findings on MRI study, it is occasionally difficult to differentiate them from PCNSL. The characteristic finding of a highly elevated lipid lactate peak on toxoplasmosis helps to stamp the lesion in question as toxoplasma. In all four of our toxoplasmosis patients, a prominent lipid lactate peak was visualised. Out of our 36 patients, only two had imaging findings consistent with HIV encephalopathy. However, conventional MRI is found to not be sensitive to brain changes in early HIV Encephalopathy[74]. MR spectroscopy is sensitive in that regard. NAA is considered the marker for viable neurons[73]. The NAA/Cr ratio in HIV Positive population is 1.6266 for all cases, 1.6695 for patients equal to or less than 40 years of age, and 1.5669 for patients more than 40 years of age. The respective values for HIV negative control group are 1.9860, 2.12 and 1.8788. On comparing the mean value of NAA/Cr to that of the normal controls we find there is a definite reduction in the value in both the age groups. In our study, spectra obtained bear this out with a reduction in the NAA peak seen in one patient, but a normal NAA peak seen in one patient. An increase in Cho/Cr has also been reported to occur in HIV encephalopathy with clear imaging findings, which was also seen in one patient. Though the number of patients is low our findings of diagnostic accuracy of 50% is in line with that of published studies[73]. Though most studies have concentrated on the utility of NAA/Cr in diagnosis of HIV Encephalopathy, **Arneth et al** found that mI/Cr is the parameter that shows earliest change in HIV encephalopathy[43]. High field (3T) MR Spectroscopy can resolve other metabolites like Glu (Glutamate) and Gln (Glutamine) which have also been found to reduce in HIV encephalopathy[75].

**Cryptococcosis**

Only 2 of our patients (5.6%) presented with cryptococcosis. The common finding of reduced NAA is seen in the spectrum in the form of reduced NAA peaks as well as moderately raised choline peaks in both the patients. This pattern signifies Trehalose peaks were not identified. Trehalose has been found in vitro studies and experimental studies[7,47] and definitive studies regarding its distribution detectable by in vivo spectroscopy were not found after an exhaustive search of the available literature.

**Stroke**

Imaging findings in stroke correlate with spectroscopic findings of reduced NAA and mild increase in lactate[10]. We had two patient of arterial thrombosis (5.6%) and two patients of dural sinus thrombosis (5.6%). The patients with arterial thrombosis showed a mild reduction in NAA but no significant lactate peak. **Pereira et al** have found that MR Spectroscopy in acute stroke may serve a prognostic purpose[76]. In patients with a lower infarct volume, a lower initial NAA concentration was associated with worse prognosis. This however did not hold in case of large infarct volume which was independently associated with worse prognosis. There were no parenchymal changes in the patients with dural venous sinus thrombosis in our study. No spectroscopic anomalies were detectable. No definitive study on the efficacy of MR Spectra in Venous sinus thrombosis was found after an exhaustive search of the available literature, although MR Spectroscopy may be used to differentiate a venous infarct from a tumour if the imaging appearance is equivocal[10]. In our study we did not have any patients of PCNSL or CMV encephalitis. A single patient of meningioma was included as he was incidentally found to be HIV positive.

Also at the time of writing, no published spectroscopy data on normal Indian population was found. The sample size for HIV negative population was small (15) but it gives an idea of the normal MR Spectroscopic findings and normal value of NAA/Cr.

## SUMMARY

The study was carried out in the Department of Radiology, KEM Hospital. A total of 50 patients were enrolled in the study, and MRI brain with MR spectroscopy was done. Out of these, 35 were HIV positive patients (cases). 15 patients with HIV negative status who also underwent MRI for unrelated pathologies were enrolled as controls. They also underwent MRI brain with MR spectroscopy. The study was done on a 1.5T Siemens Sonata Magnetom MRI Scanner.

All patients and cases underwent conventional imaging sequences (T1W, T2W, T2 FLAIR and Post contrast T1W) and Diffusion weighted imaging (DWI). Multivoxel MR Spectroscopy (MRS) with short TE (30msec) was done to determine their NAA/Cr ratio and the value from HIV positive patients was compared with data from 15 controls.

The MRI Imaging findings were reviewed and the diagnosis confirmed on clinical basis with the help of the attending physician.

Tuberculosis was the most common neurologic disease found in the HIV positive group, consisting of 9 patients. Two patients had healed granulomas. Seven of these patients had tuberculous meningitis amongst which a further 2 had vasculitic infarcts. Another two in this group had tuberculomas. Prominent lipid lactate peaks on the spectra obtained from the lesions confirmed the diagnosis of tuberculoma. PML was seen in 6 patients. NAA/Cr was found to be reduced in all the patients, and in fact the value was further reduced compared to the HIV positive group as a whole. Raised choline and myoInositol peaks were also found in all the patients.

Toxoplasmosis was seen in 4 patients. MR Spectroscopy showed lipid – lactate peaks confirming the diagnosis. A single patient with brain abscess was seen where also the spectroscopic data from the lesion confirmed the diagnosis with a lactate peak as well as amino acid resonances at characteristic locations. 2 patients had HIV encephalopathy on the imaging study. Their spectra also revealed lowered NAA peaks along with raised choline peaks.2 patients with cryptococcosis showed characteristic imaging finding of enlarged Virchow Robin (perivascular) spaces. They revealed elevated choline peaks in addition to reduced NAA. No trehalose peaks were identified. MR Spectroscopy was also done in 4 patients of vascular thrombosis, two arterial and two venous, where the MR Spectroscopic findings were not specific.

In 7 patients no focal lesions could be identified on conventional MRI sequences however on MR spectroscopic analysis NAA was found to be reduced in comparison to the controls.

The values of NAA/Cr were determined after duly processing the spectroscopic data from both cases and controls. Each group (Cases and controls) were divided on the basis of age into two age groups: Lesser than or equal to 40 years, and greater than 40 years. In all three groups the values of the mean NAA/Cr ratio was significantly reduced in comparison to controls. An ancillary finding was the reduction of NAA/Cr further in cases of PML.

Combined use of both the conventional and advanced MRI sequences is advisable as spectroscopy helps in confirming the diagnosis of opportunistic infection of the CNS in HIV positive patients. NAA/Cr ratio is reduced in HIV positive patients and is a marker for HIV infection of the brain even in the absence of imaging findings of HIV encephalopathy or when the patient is symptomatic due to neurologic disease of other etiologies.

## CONCLUSION

MRI is the optimum modality in the imaging of the brain in HIV positive patients with neurologic disease. There are a wide number of possibilities in the evaluation of HIV positive patients with neurologic disease MRI helps in early diagnosis thus helping to establish appropriate treatment.

MRI is excellent at establishing the specific disease etiology. However in specific clinical situations, MR Spectroscopy acts as a confirmatory or problem solving sequence. In the differentiation of toxoplasmosis and lymphoma, or tumor and tuberculoma, or tuberculous and pyogenic abscess, the assistance of MR Spectroscopy is required.

The use of MR Spectroscopy in establishing the diagnosis of HIV Encephalopathy in all infected patients is also elucidated. As HIV is a neurotropic virus, it affects the brain much earlier than when imaging manifestations develop. The use of MR Spectroscopy as an additional sequence in the routine evaluation of HIV patients is advisable, especially as studies have suggested that effective antiretroviral therapy may reverse the spectral changes[35] in early HIV Encephalopathy.

Conventional and advanced sequences play a complimentary role. Both need to be studied in conjunction. Use of both conventional and advanced MRI sequences is advisable to achieve a high degree of accuracy in making a complete diagnoses that is unmatched by any other investigative tool. Also MR Spectroscopy may be able to detect the onset of HIV Encephalopathy at an earlier time than conventional MRI sequences.

# REFERENCES

1. De Cock, K. M., Jaffe, H. W. & Curran, J. W. Reflections on 30 years of AIDS. *Emerg. Infect. Dis.* **17,** 1044–8 (2011).
2. Clifford, D. B. HIV-associated neurocognitive disease continues in the antiretroviral era. *Top. HIV Med.* **16,** 948
3. Valcour, V., Sithinamsuwan, P., Letendre, S. & Ances, B. Pathogenesis of HIV in the central nervous system. *Curr. HIV/AIDS Rep.* **8,** 54–61 (2011).
4. Kranick, S. M. & Nath, A. Neurologic complications of HIV-1 infection and its treatment in the era of antiretroviral therapy. *Continuum (Minneap. Minn).* **18,** 1319–37 (2012).
5. Cecchini, A. *et al.* [Neuro-AIDS: a multicenter neuroradiological study]. *Radiol. Med.* **77,** 602–12 (1989).
6. Mundinger, A. *et al.* CT and MRI: prognostic tools in patients with AIDS and neurological deficits. *Neuroradiology* **35,** 75–8 (1992).
7. Atlas, S. W. *Magnetic Resonance Imaging of the Brain and Spine, 4th edition*. (Lippincott Williams & Wilkins).
8. Tate, D. F., Khedraki, R., McCaffrey, D., Branson, D. & Dewey, J. The role of medical imaging in defining CNS abnormalities associated with HIV-infection and opportunistic infections. *Neurotherapeutics* **8,** 103–16 (2011).
9. Newton, H. B. Common neurologic complications of HIV-1 infection and AIDS. *Am. Fam. Physician* **51,** 387–98 (1995).
10. Osborn, A. G. *Diagnostic Neuroradiology*. (Elsevier, 2004).
11. Saravanan, M. & Turnbull, I. W. Brain: non-infective and non-neoplastic manifestations of HIV. *Br. J. Radiol.* **82,** 956–65 (2009).
12. Carr, H. Free precession techniques in nuclear magnetic resonance. (1953).
13. Damadian, R. Tumor detection by nuclear magnetic resonance. *Science* **171,** 1151–3 (1971).
14. Van der Graaf, M. In vivo magnetic resonance spectroscopy: basic methodology and clinical applications. *Eur. Biophys. J.* **39,** 527–40 (2010).
15. Schmidt, C. W. Metabolomics: what's happening downstream of DNA. *Environ. Health Perspect.* **112,** A410–5 (2004).
16. Bertholdo, D., Watcharakorn, A. & Castillo, M. Brain proton magnetic resonance spectroscopy: introduction and overview. *Neuroimaging Clin. N. Am.* **23,** 359–80 (2013).
17. Longo DL, Kasper DL, Jameson JL, Fauci AS, Hauser SL, L. J. *Harrison's Principles of Internal Medicine*. (McGraw Hill).
18. Simpson, D. M. Human immunodeficiency virus-associated dementia: review of pathogenesis, prophylaxis, and treatment studies of zidovudine therapy. *Clin. Infect. Dis.* **29,** 19–34 (1999).
19. Martin-Duverneuil, N. *et al.* [Cerebral toxoplasmosis. Neuroradiologic diagnosis and prognostic monitoring]. *J. Neuroradiol.* **22,** 196–203 (1995).
20. Kumar V, Abbas AK, Fausto N, A. J. *Robbins and Cotran - The Pathologic Basis of Disease*. (Saunders Elsevier, 2010).
21. Senocak, E. *et al.* Imaging features of CNS involvement in AIDS. *Diagn. Interv. Radiol.* **16,** 193–200 (2010).
22. Haldorsen, I. S., Espeland, A. & Larsson, E.-M. Central nervous system lymphoma: characteristic findings on traditional and advanced imaging. *AJNR. Am. J. Neuroradiol.* **32,** 984–92 (2011).
23. Hochberg, F. H. & Miller, D. C. Primary central nervous system lymphoma. *J. Neurosurg.* **68,** 835–53 (1988).
24. Manenti, G. *et al.* A Case of Primary T-Cell Central Nervous System Lymphoma: MR Imaging and MR Spectroscopy Assessment. *Case Rep. Radiol.* **2013,** 916348 (2013).
25. Zhang, D. *et al.* MRI findings of primary CNS lymphoma in 26 immunocompetent patients. *Korean J. Radiol.* **11,** 269–77
26. Lolli, V., Tampieri, D., Melançon, D. & Delpilar Cortes, M. Imaging in primary central nervous system lymphoma. *Neuroradiol. J.* **23,** 680–9 (2010).
27. Berger, J. R., Levy, R. M., Flomenhoft, D. & Dobbs, M. Predictive factors for prolonged survival in acquired immunodeficiency syndrome-associated progressive multifocal leukoencephalopathy. *Ann. Neurol.* **44,** 341–9 (1998).
28. Giancola, M. L. *et al.* Progressive multifocal leukoencephalopathy in HIV-infected patients in the era of HAART: radiological features at diagnosis and follow-up and correlation with clinical variables. *AIDS Res. Hum. Retroviruses* **24,** 155–62 (2008).

29. Yoon, J. H., Bang, O. Y. & Kim, H. S. Progressive Multifocal Leukoencephalopathy in AIDS: Proton MR Spectroscopy Patterns of Asynchronous Lesions Confirmed by Serial Diffusion-Weighted Imaging and Apparent Diffusion Coefficient Mapping. *J. Clin. Neurol.* **3,** 200–3 (2007).
30. Chang, L. *et al.* Metabolite abnormalities in progressive multifocal leukoencephalopathy by proton magnetic resonance spectroscopy. *Neurology* **48,** 836–45 (1997).
31. Gheuens, S. *et al.* Metabolic profile of PML lesions in patients with and without IRIS: an observational study. *Neurology* **79,** 1041–8 (2012).
32. Bag, A. K., Curé, J. K., Chapman, P. R., Roberson, G. H. & Shah, R. JC virus infection of the brain. *AJNR. Am. J. Neuroradiol.* **31,** 1564–76 (2010).
33. Walot, I., Miller, B. L., Chang, L. & Mehringer, C. M. Neuroimaging findings in patients with AIDS. *Clin. Infect. Dis.* **22,** 906–19 (1996).
34. Dousset, V. *et al.* Magnetization transfer study of HIV encephalitis and progressive multifocal leukoencephalopathy. Groupe d'Epidémiologie Clinique du SIDA en Aquitaine. *AJNR. Am. J. Neuroradiol.* **18,** 895–901 (1997).
35. Chang, L. *et al.* Highly active antiretroviral therapy reverses brain metabolite abnormalities in mild HIV dementia. *Neurology* **53,** 782–9 (1999).
36. Sailasuta, N. *et al.* Change in brain magnetic resonance spectroscopy after treatment during acute HIV infection. *PLoS One* **7,** e49272 (2012).
37. Ratai, E.-M. *et al.* In vivo proton magnetic resonance spectroscopy reveals region specific metabolic responses to SIV infection in the macaque brain. *BMC Neurosci.* **10,** 63 (2009).
38. Lentz, M. R. *et al.* Changes in MRS neuronal markers and T cell phenotypes observed during early HIV infection. *Neurology* **72,** 1465–72 (2009).
39. Ernst, T., Jiang, C. S., Nakama, H., Buchthal, S. & Chang, L. Lower brain glutamate is associated with cognitive deficits in HIV patients: a new mechanism for HIV-associated neurocognitive disorder. *J. Magn. Reson. Imaging* **32,** 1045–53 (2010).
40. Tarasów, E. *et al.* Cerebral MR spectroscopy in neurologically asymptomatic HIV-infected patients. *Acta Radiol.* **44,** 206–12 (2003).
41. Chang, L. *et al.* A multicenter in vivo proton-MRS study of HIV-associated dementia and its relationship to age. *Neuroimage* **23,** 1336–47 (2004).
42. Suwanwelaa, N. *et al.* Magnetic resonance spectroscopy of the brain in neurologically asymptomatic HIV-infected patients. *Magn. Reson. Imaging* **18,** 859–65 (2000).
43. Arneth, B., Pilatus, U., Lanfermann, H. & Enzensberger, W. Objectification and quantification of the cognitive impairment from an existing HIV infection or HIV encephalopathy using magnetic resonance spectroscopy. *J. Int. Assoc. Provid. AIDS Care* **12,** 253–60
44. Corti, M., Villafañe, M. F., Negroni, R., Arechavala, A. & Maiolo, E. Magnetic resonance imaging findings in AIDS patients with central nervous system cryptococcosis. *Rev. Iberoam. Micol.* **25,** 211–4 (2008).
45. Tien, R. D., Chu, P. K., Hesselink, J. R., Duberg, A. & Wiley, C. Intracranial cryptococcosis in immunocompromised patients: CT and MR findings in 29 cases. *AJNR. Am. J. Neuroradiol.* **12,** 283–9
46. Luthra, G. *et al.* Comparative evaluation of fungal, tubercular, and pyogenic brain abscesses with conventional and diffusion MR imaging and proton MR spectroscopy. *AJNR. Am. J. Neuroradiol.* **28,** 1332–8 (2007).
47. Himmelreich, U. *et al.* Cryptococcomas distinguished from gliomas with MR spectroscopy: an experimental rat and cell culture study. *Radiology* **220,** 122–8 (2001).
48. Cherian, A. & Thomas, S. V. Central nervous system tuberculosis. *Afr. Health Sci.* **11,** 116–27 (2011).
49. Corbett, E. L. *et al.* The growing burden of tuberculosis: global trends and interactions with the HIV epidemic. *Arch. Intern. Med.* **163,** 1009–21 (2003).
50. Harries, A. D., Zachariah, R. & Lawn, S. D. Providing HIV care for co-infected tuberculosis patients: a perspective from sub-Saharan Africa. *Int. J. Tuberc. Lung Dis.* **13,** 6–16 (2009).
51. Garg, R. K. Tuberculosis of the central nervous system. *Postgrad. Med. J.* **75,** 133–40 (1999).
52. Sengöz, G. [Evaluating 82 cases of tuberculous meningitis]. *Tuberk. Toraks* **53,** 51–6 (2005).
53. Trivedi, R., Saksena, S. & Gupta, R. K. Magnetic resonance imaging in central nervous system tuberculosis. *Indian J. Radiol. Imaging* **19,** 256–65 (2009).
54. Nelson, C. A. & Zunt, J. R. Tuberculosis of the central nervous system in immunocompromised patients: HIV infection and solid organ transplant recipients. *Clin. Infect. Dis.* **53,** 915–26 (2011).
55. Singh, I. *et al.* Role of endoscopic third ventriculostomy in patients with communicating hydrocephalus: an evaluation by MR ventriculography. *Neurosurg. Rev.* **31,** 319–25 (2008).
56. Gupta, R. K., Kathuria, M. K. & Pradhan, S. Magnetization transfer MR imaging in CNS tuberculosis. *AJNR. Am. J. Neuroradiol.* **20,** 867–75 (1999).
57. Kim, H.-J. *et al.* Tuberculous encephalopathy without meningitis: pathology and brain MRI findings. *Eur. Neurol.* **65,** 156–9 (2011).
58. Gupta, R. K. *et al.* MR imaging and angiography in tuberculous meningitis. *Neuroradiology* **36,** 87–92 (1994).

59. Gupta, R. K. & Kumar, S. Central nervous system tuberculosis. *Neuroimaging Clin. N. Am.* **21,** 795–814, vii–viii (2011).
60. Lu, M. Imaging diagnosis of spinal intramedullary tuberculoma: case reports and literature review. *J. Spinal Cord Med.* **33,** 159–62 (2010).
61. Whitener, D. R. Tuberculous brain abscess. Report of a case and review of the literature. *Arch. Neurol.* **35,** 148–55 (1978).
62. Erdoğan, E. & Cansever, T. Pyogenic brain abscess. *Neurosurg. Focus* **24,** E2 (2008).
63. Pal, D. *et al.* In vivo proton MR spectroscopy evaluation of pyogenic brain abscesses: a report of 194 cases. *AJNR. Am. J. Neuroradiol.* **31,** 360–6 (2010).
64. Nagel, M. A., Mahalingam, R., Cohrs, R. J. & Gilden, D. Virus vasculopathy and stroke: an under-recognized cause and treatment target. *Infect. Disord. Drug Targets* **10,** 105–11 (2010).
65. Benjamin, L. A. *et al.* HIV infection and stroke: current perspectives and future directions. *Lancet Neurol.* **11,** 878–90 (2012).
66. Ramadan, N. M., Deveshwar, R. & Levine, S. R. Magnetic resonance and clinical cerebrovascular disease. An update. *Stroke* **20,** 1279–1283 (1989).
67. Dani, K. A., An, L., Henning, E. C., Shen, J. & Warach, S. Multivoxel MR spectroscopy in acute ischemic stroke: comparison to the stroke protocol MRI. *Stroke.* **43,** 2962–7 (2012).
68. Collazos, J. Opportunistic infections of the CNS in patients with AIDS: diagnosis and management. *CNS Drugs* **17,** 869–87 (2003).
69. Luma, H. N. *et al.* HIV-Associated Central Nervous System Disease in Patients Admitted at the Douala General Hospital between 2004 and 2009: A Retrospective Study. *AIDS Res. Treat.* **2013,** 709810 (2013).
70. Kongsiriwattanakul, S. & Suankratay, C. Central nervous system infections in HIV-infected patients hospitalized at King Chulalongkorn Memorial Hospital. *J. Med. Assoc. Thai.* **94,** 551–8 (2011).
71. Choe, P. G. *et al.* Spectrum of intracranial parenchymal lesions in patients with human immunodeficiency virus infection in the Republic of Korea. *J. Korean Med. Sci.* **25,** 1005–10 (2010).
72. Lanjewar, D. N., Jain, P. P. & Shetty, C. R. Profile of central nervous system pathology in patients with AIDS: an autopsy study from India. *AIDS* **12,** 309–13 (1998).
73. Simone, I. L. *et al.* Localised 1H-MR spectroscopy for metabolic characterisation of diffuse and focal brain lesions in patients infected with HIV. *J. Neurol. Neurosurg. Psychiatry* **64,** 516–23 (1998).
74. Prado, P. T. C., Escorsi-Rosset, S., Cervi, M. C. & Santos, A. C. Image evaluation of HIV encephalopathy: a multimodal approach using quantitative MR techniques. *Neuroradiology* **53,** 899–908 (2011).
75. Mohamed, M. A. *et al.* Brain metabolism and cognitive impairment in HIV infection: a 3-T magnetic resonance spectroscopy study. *Magn. Reson. Imaging* **28,** 1251–7 (2010).
76. Pereira, A. C. *et al.* Measurement of initial N-acetyl aspartate concentration by magnetic resonance spectroscopy and initial infarct volume by MRI predicts outcome in patients with middle cerebral artery territory infarction. *Stroke.* **30,** 1577–82 (1999).

# ANNEXURE

## ANNEXURE – A : ABBREVIATIONS

**CT-** Computed Tomography

**MRI -** Magnetic Resonance Imaging

**MRS –** Magnetic Resonance Spectroscopy

**NMR –** Nuclear Magnetic Resonance

**SVS –** Single Voxel Spectroscopy

**CSI –** Chemical Shift Imaging

**SNR –** Signal to noise Ratio

**MTR –** Magnetization transfer ratio

**TE -** Time to Echo

**TR-** Time to Repeat

**T1W-** T1 Weighted

**T2WI-** T2 Weighted

**GRE-** Gradient Recalled Echo

**FLAIR –** Fluid Attenuated Inversion Recovery

**DWI –** Diffusion Weighted Imaging

**PWI –** Perfusion Weighted Imaging

**SWI –** Susceptibility Weighted Imaging

**$^1$H -** Proton

**NAA -** *N*-acetylaspartate

**Cho –** Choline

**Cr –** Creatine and phosphocreatine

**Lac –** Lactate

**Glx –** Glutamate and Glutamine

**Gln –** Glutamine

**mI –** Myoinositol

**HIV –** Human Immunodeficiency Virus

**AIDS**: Acquired Immunodeficiency Syndrome

**IRIS -** Immune reconstitution inflammatory syndrome

**PML-** Progressive Multifocal Leukoencephalopathy

**CVD –** Cerebrovascular Disease

**PCNSL –** Primary Central Nervous system Lymphoma

**CMV -** Cytomegalovirus

**TBM –** Tuberculous meningitis

**SSS –** Superior sagittal sinus

**MCA -** Middle cerebral artery

**CNS –** Central Nervous system

**HAART -** Highly active antiretroviral therapy

**ART -** Antiretroviral therapy

# ANNEXURE – C : CASE RECORD FORM

Name:                                                                Date:

Age:

Sex :    Male/Female

Indoor patient number:                                     Outdoor patient number:

Date of Examination

Number:

Referring physician

Weight of the patient:

Chief complaints                                              Duration of symptoms :

Any other complaints:

Past H/O:

Relevant neurological examination findings:

Current CD4 count (for HIV positive patients only):

MR Spectroscopy Findings:

    i)     NAA/Cr Ratio:

    ii)    Analysis Of Spectroscopy Curve:

Final diagnosis:

Signature of the Chief Investigator